\documentclass[aps,prd,twocolumn,groupedaddress]{revtex4-1}

\usepackage{graphicx}
\usepackage{subfigure}
\usepackage{amsmath} 

\usepackage[colorlinks=true,citecolor=blue,urlcolor=magenta,breaklinks]{hyperref}
\usepackage{natbib}
\usepackage[T1]{fontenc}
\usepackage{times}
\newcommand{\Lagr}{\mathcal{L}} 
\newcommand{\Hami}{\mathcal{H}}  

\begin{document}


%
\title{New neutrino physics and the altered shapes of solar  neutrino spectra}
\author{Il\'idio Lopes}
\email[]{ilidio.lopes@tecnico.ulisboa.pt}
\affiliation{Centro Multidisciplinar de Astrof\'{\i}sica, Instituto Superior T\'ecnico, 
Universidade de Lisboa , Av. Rovisco Pais, 1049-001 Lisboa, Portugal}
%
%
%

\begin{abstract}
Neutrinos coming from the Sun's core are now measured with a high precision,
and fundamental neutrino oscillations parameters are determined with a good accuracy. 
In this work, we estimate the impact that a new neutrino physics model, the so-called generalized Mikheyev-Smirnov-Wolfenstein (MSW)  
oscillation mechanism, has on the shape of some of leading solar neutrino spectra, some of which will be partially tested by the next generation 
of solar neutrino experiments.  In these calculations, we use a high-precision standard solar model
in good agreement with helioseismology data.
We found that the neutrino spectra of the different solar nuclear reactions of the proton-proton  chains and carbon-nitrogen-oxygen   
cycle have quite distinct sensitivities to the new neutrino physics. The $HeP$ and $^8B$ neutrino 
spectra are the ones for which their shapes are more affected when neutrinos 
interact with quarks in addition to electrons.  
The shape of the $^{15}O$ and $^{17}F$ neutrino spectra are also modified, although in these cases 
the impact is much smaller. Finally, the impact in the shape of the $PP$ and $^{13}N$ neutrino spectra is practically negligible.       
\end{abstract}


\maketitle


\section{Introduction\label{sec-intro}}

Since their discovery in 1956, neutrinos have always surprised physicists due to their unexpected properties, 
often challenging our basic understanding of the standard model of particle physics
(in the remainder of the article, it will be called simply the 'standard model')
and the properties of elementary particles.   
In particular, the discovery of neutrino flavour oscillations stands as one of the most convincing  
proofs that the standard model   
is incomplete as it does not explain all the known  
experimental properties of the fundamental particles. 

\smallskip
The neutrino research success has been made possible mostly due to many dedicated experiments performed during the last fifty years. It is worth highlighting the contributions of some pioneering experiments, among others, such as the Super-Kamiokande detector~\citep{2001ARNPS..51..451J,2001RvMP...73...85K} where the oscillation of atmospheric neutrinos was discovered, and the SNO detector~\citep{2001PhRvL..87g1301A} where  the fluxes of all neutrino flavour species 
produced in the Sun's core were measured for the first time. Many other experiments done during the previous decades have contributed for the success of this story, in particular, the solar neutrino experiments. Despite their technical  complexities, these experiments were able 
to measure the electron neutrino fluxes coming from the Sun and played a major role 
in the establishment of the so-called  solar neutrino problem -- a  discrepancy  
between the theoretical prediction of neutrino fluxes and their experimental measurements: 
the experimental value being one third of the predicted value. 
This fact was  evidenced for the first time by the Homestake Experiment of Ray Davis~\citep{1968PhRvL..20.1205D},
and confirmed by many other experiments that followed. It was the solar neutrino problem that prompted the development of  
the neutrino flavour oscillation model.

\smallskip
If indeed the previous generation of  solar neutrino detectors has been one of the beacons of particle physics, both by leading the way in uncovering the basic properties of particles, including the nature of neutrino flavour oscillations,  and by being responsible for developing pioneering techniques in experimental neutrino detection~\citep{2013ARAA..51...21H}, the next generation of detectors 
is equally promising in discovering new physics. Among various, some of which
will be looking for evidence of neutrino new physics we can mention the following future detectors:  
the Low Energy Neutrino Astronomy~\citep[LENA,][]{2012APh....35..685W},
the Jiangmen Underground Neutrino Observatory~\citep[JUNO,][]{2016JPhG...43c0401A},
the Deep Underground Neutrino Experiment~\citep[DUNE,][]{2016NuPhB.908..318D},
the NO$\nu$A Neutrino Experiment~\citep[NO$\nu$A,][]{2012arXiv1207.6642F}, and 
the Jinping Neutrino Experiment~\citep[Jinping,][]{2016arXiv160201733B}.
 
These detectors will measure with high precision the neutrino fluxes 
and neutrino spectra of a few key neutrino nuclear reactions, such as the  
$^8$B electron-neutrino ($^8$B$\nu_e$) spectrum produced by the $\beta$-decay process in the $^8$B solar (chain) 
reaction: $^7Be(p,\gamma)^8(e^{+}\nu_e)^8B^*(\alpha)^4He$~\citep{2009RPPh...72j6201B,2013arXiv1310.4340D}.  
This will allow us to probe in detail the Sun's core, including the search for new neutrino physics interaction
or even new physics processes.  Moreover, the high quality of the data will enable the  development of  
inversion techniques for determining  basic properties of the solar plasma~\citep[e.g.,][]{2005NuPhS.138..347B}. 
Specific examples can be found in~\citet{1998PhLB..427..317B} and~\citet{2013ApJ...777L...7L}. 
Equally, solar neutrino data can be used to find specific features associated with possible new physical 
processes present in the Sun's interior~\citep[e.g.][]{2013PhRvD..87d3001S}, 
such as the possibility of an isothermal solar core associated with the presence of
dark matter~\citep{2002PhRvL..88o1303L}.

\smallskip
Today, the basic principles of neutrino physics are firmly established, neutrinos  
are massive particles with a lepton flavours mix. The parameters describing neutrino flavour oscillations 
are measured with great accuracy and precision, which has been possible due to the extensive studies
made by many different types of neutrino experiments: solar and atmospheric neutrino observatories, nuclear reactors  
and experimental particle accelerators~\citep[e.g.,][]{2010PhRvD..81i2004W,2013PhRvL.110q1801A,2014PhRvL.112f1802A}. 
Section~\ref{sb-nodp} presents the status of the current neutrino oscillation parameters obtained 
from up-to-date experimental data. 

\medskip

Even if many properties of neutrinos are known, many others are still a mystery:

- firstly, are neutrinos Majorana or Dirac fermions ? i.e., are neutrinos  their own anti-particle ? 
Although the theoretical expectation favours the first option, only  experimental evidence can settle this question;
 
- secondly, what is the mass hierarchy of neutrinos ? In other words, does the order of neutrino masses 
between the different particle families follow a {\it normal hierarchy} 
-- two light neutrinos followed by a heavier one,  or an {\it inverted hierarchy} 
-- one light neutrino follow by two heavier ones ?

\smallskip

Together with the CP violation in the lepton sector, 
these are the most important questions  of neutrino physics.
Some of these questions will be answered by the next generation of neutrino experiments -- 
the long baseline neutrino experiments and solar neutrino telescopes. Nevertheless, 
it is necessary to improve the current  neutrino flavour model to take full  advantage 
of the forthcoming experimental data.     
  
\smallskip
Despite the success of the current neutrino physics model in explaining most of the neutrino's known observed properties, the solution encountered clearly indicate the existence of new physics beyond the standard model. 
As such, this implies that within the current particle physics  theoretical framework, experiments can study neutrinos in other types of interactions.
When such processes occur, these lead to important modifications of the physical mechanisms by which neutrinos are created, propagate and interact with other particles of the standard model. This new class of neutrino interactions is usually known as non-standard interactions ($nsi$). 

\smallskip
The  non-standard interactions of neutrinos have been extensively studied in the literature, among others reviews on this topic,
see for instance~\citet[][]{2015NJPh...17i5002M,2013RPPh...76d4201O}. Moreover, the constraints on the $nsi$ parameters and their effects for low energy neutrinos have been derived from a great variety of experimental results. Until now no definitive evidence of non-standard interactions has been provided by the experimental data. 
Actually, all observations made as yet can be explained in terms of the standard interactions of the three known neutrinos, although some of them need the help of sterile neutrinos. Nevertheless, in some cases the non-standard interactions of neutrinos
provide an interesting and valid alternative~\citep[e.g.,][]{2014NuPhB.886...31G}

\smallskip
In this work we are mostly concerned with the non-standard interactions of solar neutrinos. These interactions can affect the neutrino production inside the Sun, the detection of neutrinos by experimental detectors and the neutrino propagation in the Earth's and Sun's interiors. In particular, our study focus on the propagation of neutrinos through baryonic matter in the Sun's 
interior, a process usually known as the generalized Mikheyev-Smirnov-Wolfenstein ($MSW$) oscillation mechanism, or generalized matter effect oscillations. Our goal is to make predictions about the modifications imprinted by this new generalized $MSW$ on the shape of the solar neutrino spectrum produced by some of the (pp) and carbon-nitrogen-oxygen (CNO) key nuclear reactions, like $HeP$ and $^8B$ neutrino spectra.   

\smallskip
The high quality of the standard solar model in reproducing the measured solar neutrino fluxes, 
and the observed acoustic frequency oscillations, make it a privileged tool to look for the new interactions
within a generalized Mikheyev-Smirnov-Wolfenstein mechanism occurring in the Sun's interior. 
The standard solar model~\citep[SSM,][]{1993ApJ...408..347T} partly validated by helioseismology,
predicts that the density inside the Sun varies from  about 150 ${\rm g\;cm^{-3}}$ in the centre of the star, 
to  1 ${\rm g\;cm^{-3} }$ at half of the solar radius. The variation of density of matter with the solar radius
is followed by identical variations on the local quantities of electrons and quarks. Moreover, the different type
of quarks will also be affected by the local distribution of chemical elements (most noticeably Hydrogen and Helium)    
which lead to a not obvious distribution of up- and down-quarks.  Therefore, we can anticipate that 
the current standard solar model combined  with data coming from the next generation of the solar neutrino detections, 
 will allow us to put much stronger constraints in the non-standard interactions of neutrinos.    

In the next section, we review the current status of the standard solar model and neutrino production
in the Sun's core, In Section~\ref{sec-MNPO}, we present a summarised discussion   about the current standard  
neutrino oscillation flavour model, and a generalized model for which neutrinos have new types 
of interactions with standard particles. In Section~\ref{sec-SEMNS}, we compute the neutrino
spectra resulting from these new types of interactions. In the final section we 
discuss the results and their implications for the future neutrino  experiments.

\section{Neutrino production in the Sun's core}
\label{sec-NPSC}
\begin{figure}[!t]
\centering 
\includegraphics[scale=0.45]{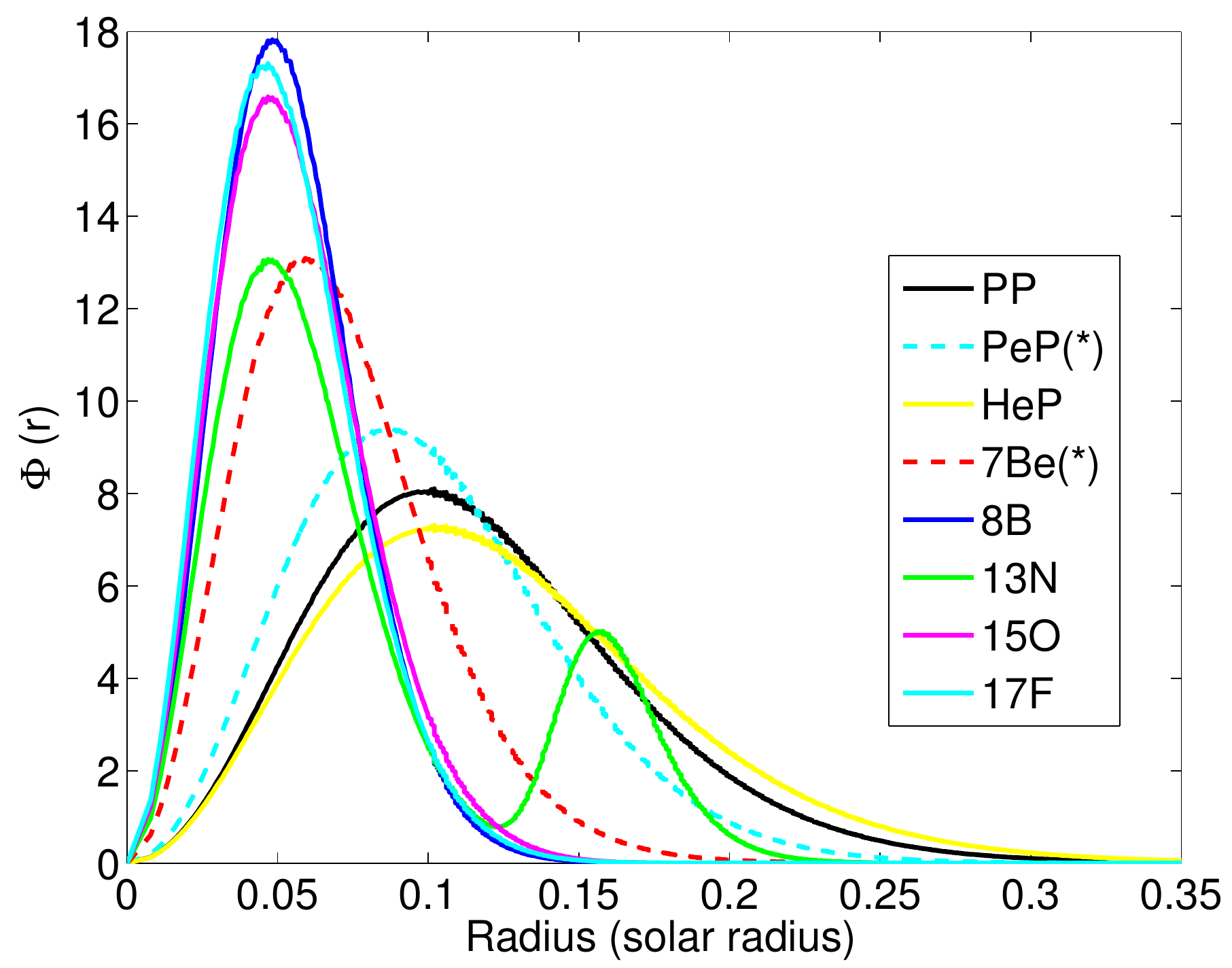}
\caption{The electron-neutrino fluxes produced in the various nuclear reactions of the  
pp chains and CNO cycle.These neutrino fluxes were calculated for a standard solar model 
using the most updated microscopic physics data.
This solar model is in agreement with the most current  helioseismology diagnostic and other solar standard models 
published in the literature (see text).
For each neutrino type $j$ (with $j=PP,PeP(*),HeP,^8B,^7Be(*),^{13}N,^{15}O,^{17}F$),
$\Phi_j (r)\equiv (1/F_j)\;d{f_j(r)}/dr$ is drawn
as a function of the fractional radius $r$ for which $f_j$ is the flux in ${\rm s^{-1}}$ 
and $F_j$ is the total flux for this neutrino type. The neutrino sources noted with the symbol $(*)$ 
correspond to spectral lines. The same colour scheme is used in figures~\ref{figure:4} and~\ref{figure:5}.
}
\label{figure:1}
\end{figure}

\subsection{Helioseismology and the standard solar model}

During the last three decades helioseismology has provided solar physics with a tool
that describes with unprecedented quality the internal structure of the Sun from its surface 
up to the deepest layers of the Sun's interior. This has allowed astronomers to characterise with great
precision the different solar neutrino sources. Equally, this discipline has
stimulated the development of inversion techniques to probe the internal solar dynamics.
Today an impressive agreement   has been reached  between the neutrino flux predictions 
and  the neutrino flux measurements made by the existing neutrino detectors.
The high quality of the helioseismology data has allowed to compute an exceptionally accurate model
of the Sun's interior - the standard solar model. The neutrino fluxes predictions of the 
solar model have an accuracy comparable to the current measurements made by particle accelerators or nuclear reactors.

\smallskip
The standard solar model in this study is obtained using a version of the one-dimensional 
stellar evolution code CESAM~\citep{1997AAS..124..597M}.  
The code has an up-to-date and very refined microscopic physics
(updated equation of state, opacities, nuclear reactions rates, 
and an accurate treatment of the microscopic diffusion of heavy elements), 
including the solar mixture of~\citet{2005ASPC..336...25A,2009ARAA..47..481A}.
This solar model is calibrated to reproduce with high accuracy the present total radius,
luminosity and mass of the Sun at the present  $t_\odot = 4.54\pm 0.04\; {\rm Gyr}$~\citep{2011RPPh...74h6901T}. 
Moreover, this model is required to have a fixed value of the photospheric ratio $(Z/X)_\odot$, where  
X and Z are the mass fraction of hydrogen and the mass fraction of elements heavier than helium.
This solar standard model  shows acoustic seismic diagnostics 
and solar neutrino fluxes similar to other models found in the literature~ 
\citep{1993ApJ...408..347T,2004PhRvL..93u1102T,2005ApJ...618.1049B,2005ApJ...621L..85B,2009ApJ...705L.123S,2010ApJ...713.1108G,2010ApJ...715.1539T}

\smallskip
This solar model is calibrated for the present day solar data with a high accuracy. 
Therefore slightly different physical assumptions, 
will lead to different radial profiles of temperature, density  and chemical composition,
among other quantities. These changes result from readjustments of the Sun's internal structure 
caused by the need to obtain the same total luminosity. In particular, the  neutrino fluxes and 
sound speed profile will be very sensitive to the radial distributions of the previous quantities.
As such, using the high precision data from helioseismology, it is possible to put strong 
constraints to the internal structure of the Sun and its  neutrino fluxes~\citep{2011RPPh...74h6901T,2012RAA....12.1107T}. 

\smallskip
The current uncertainty  between the square of the sound speed profile inferred from helioseismology acoustic data
and the one obtained from the standard solar model using the up-to-date photospheric abundances~\citet{2009ARAA..47..481A} is smaller than 3\% for any layer of the Sun's interior. Although  there is a difference  between the sound speed profile computed  using an older mixture of abundances by~\citet{1998SSRv...85..161G}  or the new mixture of~\citet{2009ARAA..47..481A}, for this study these effects are negligible on the radial variation of electrons, protons and neutrons. 
This is even more so, since recent measurements of the solar metallicity abundances suggest that the sound speed difference of helioseismic data and standard solar model is reduced further~\citep{2016arXiv160305960V}.
 
\smallskip
Particularly relevant for our study is the radial profile of the electron, proton and neutron densities 
inside the Sun, since these quantities are  fundamental  ingredients  to test the non-standard neutrino physics theories.

\begin{figure}[!t]
\centering 
\includegraphics[scale=0.45]{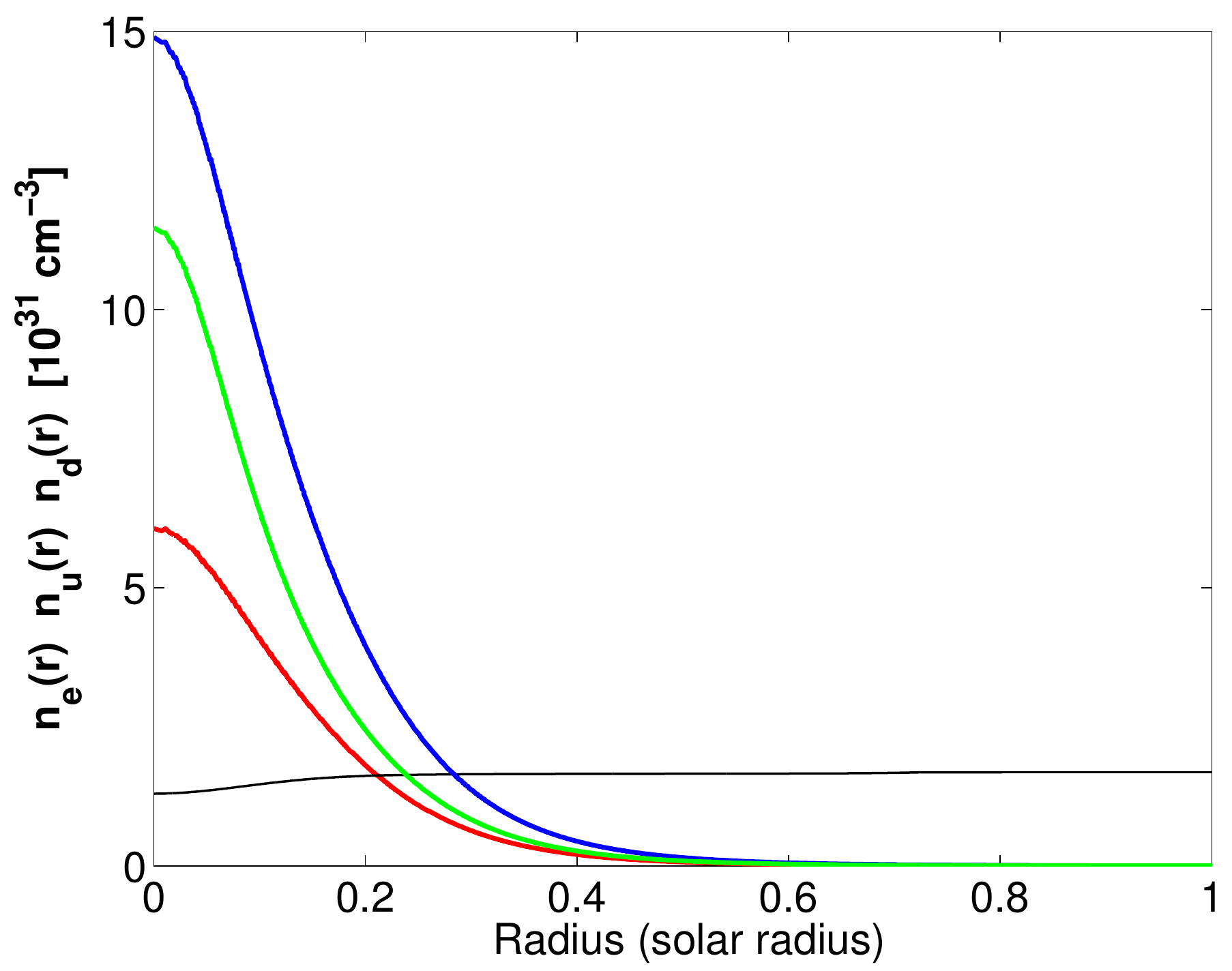}
\caption{The solar plasma is constituted mostly by electrons, protons and neutrons. 
Variation of the number density of electrons $ n_e (r) $ (red curve), up-quarks $ n_u (r) $ (blue curve) and 
down-quarks $ n_d (r) $ (green curve), and the relative variation of 
up and down quarks $n_u(r)/n_d(r)$ (black curve). In the center $n_u/n_d=1.3$ and near the surface $n_u/n_d=1.7$.
}
\label{figure:2}
\end{figure}

\subsection{The solar neutrino sources}

The neutrino fluxes produced in the nuclear reactions of the pp chains and CNO cycle 
have been computed for an updated version of the solar standard model, as discussed in the previous section.  
Figure~\ref{figure:1} shows the location of the different neutrino emission regions of the nuclear reactions
for an up-to-date SSM. In the Sun's core, the neutrino emission regions occur in a sequence of shells, following closely  
the location of nuclear reactions, orderly arranged in a sequence dependent on their temperature.  
The helioseismology data and solar neutrino fluxes guaranties that such neutrino shells  are known with a great accuracy. 
The leading source of the energy in the present Sun are the pp chains nuclear reactions, since the  CNO cycle 
nuclear reactions contribute with less than 2\%.  The first reaction of the pp chains
is the $PP$-$\nu$ reaction which has the largest neutrino emission shell, 
a region that extends from the centre  to 0.30 $R_\odot$. The $PeP$-$\nu$ reaction has 
a neutrino emission shell which is similar to the $PP$ reaction, but with a shell of 0.25 $R_\odot$.  
These nuclear reactions are strongly dependent on the total luminosity of the star.
Alternatively, the neutrino emission shells of $^8B$-$\nu$ and $^7Be$-$\nu$ extend up to  0.15 $R_\odot$ and 0.22$R_\odot$. 
It is interesting to notice that the maximum emission of neutrinos for the pp chains 
nuclear reactions, follows an ordered sequence (see figure~\ref{figure:1}): $^8B$-$\nu$, $^7Be$-$\nu$, $PeP$-$\nu $ and $PP$-$\nu$  
with the maximum emission located at 0.05, 0.06, 0.08 and 0.10 $R_\odot$. 
The known neutrino emission shells of the different CNO cycle nuclear reactions are the following ones:
$^{15}O$-$\nu$, $^{17}F$-$\nu$ and $^{13}N$-$\nu$. These shells are  similar to the  
 $^8B$-$\nu$ emission shell. The $^{13}N$-$\nu $  have two independent shells: one in the Sun's deepest layers of the 
core and  a second  shell located between 0.12 and 0.25 of $R_\odot$. The emission  of neutrinos 
for $^{15}O$-$\nu$, $^{17}F$-$\nu$ and $^{13}N$-$\nu$ shells is maximal at 0.04 - 0.05 of  $R_\odot$. 
The $^{13}N$-$\nu$ neutrinos have a second emission maximum which is located at 0.16 $R_\odot$.

\subsection{Neutrinos, Electrons and Quarks}

The electron density  $ n_e (r) =N_{o}\; \rho(r)/\mu_e(r) $ 
where $\mu_e$ is the mean molecular weight per electron, $\rho(r)$ the density of matter 
and $N_{o}$ the Avogadro's number.  In this model we will consider the impact on the 
quarks up and down.  Accordingly, the density of up and down quarks  
will be computed  from a relation analogous to $ n_e (r)$,  $ n_i (r) =N_{o}\; \rho(r)/\mu_i(r) $ 
(with $i=u,d$) where $\mu_i(r)$ is the mean molecular weight per quark
given by      
\begin{eqnarray}
\mu_i(r)=\left[(1+\delta_{iu}) X(r)+\frac{3}{2} Y(r)+\frac{3}{2} Z (r) \right]^{-1}
\end{eqnarray} 
where $i=u,d$ with $X+Y+Z=1$. The distribution of electrons and up and down quarks, as a function
of the radius of the Sun for the standard solar model, is shown in figure~\ref{figure:2}.
The mean molecular weight per quark is dominated by Hydrogen and Helium since the only
other elements included in $Z$ like Carbon, Nitrogen, Oxygen and heavier elements contribute with a very
small fraction to the solar plasma. Although the Z-variation can affect the evolution of the
star in the way it affects the radiative transport~\citep{2016arXiv160305960V}, 
its impact on the $si$ and $nsi$ MSW interactions for $\mu_i(r),\; i=u,d$) is small
since the relative radial variation between up- and down-quarks due to Z-variation is not significant.

\section{Model of neutrino physics oscillations}
\label{sec-MNPO}
\subsection{basic neutrino physics}
In the Standard Model, neutrinos interact with other particles only via weak standard interactions $(si)$, 
which are described by the Lagragian $\Lagr_{si}$ which can be decomposed into components describing
the charged and neutral interactions~\citep{2008PhR...460....1G,2013arXiv1310.7858B,2015arXiv150202928P}.
Nevertheless, in the current study,  we choose to write the Lagragian $\Lagr_{si}$ 
as an effective interaction Lagrangian~\citep{2003JHEP...03..011D,2015arXiv150202928P}, 
which at low and intermediate neutrino energies reads
\begin{eqnarray}
\Lagr_{si}=-2\sqrt{2}G_F
g_p^f 
\left( \bar{\nu}_\alpha\gamma_\rho L\nu_\alpha \right)
\left( \bar{f}\gamma^\rho P f \right)
\label{eq:Lagrsi}
\end{eqnarray}
where $f$ denotes a lepton or a quark, such as  the $u-$quark and the $d-$quark, $\nu_\alpha$ are the three light neutrinos (with the subscript $\alpha=e,\mu,\tau$),  
$P$ is the chiral projector (is equal to $R$ or $L$ such that $R,L\equiv (1\pm \gamma^5)/2$) 
and  $g_p^f$ denotes the strength of the interaction ($ns$) as defined in the standard model  between neutrinos of flavours $\alpha$ and $\beta$ and the $P$-handed component 
of the fermion $f$. Specifically, the $g_p^f$ coupling (left- and right- handed coupling, i.e., $g_L^f$ and $g_R^f$) 
for the  $u$-quark (and $c$- and $t$-quark) to the $Z-$boson corresponds to $g^u_L=1/2-2/3\sin^2{\theta}_w$ and $g^u_R=-2/3\sin^2{\theta}_w$.
Similarly, the $g_p^f$ for the $d$-quark (and $s$- and $b$-quark) corresponds to
$g^d_L=-1/2+1/3\sin^2{\theta}_w$ and $g^d_R=1/3\sin^2{\theta}_w$, 
the  $g_p^f$ for the electron corresponds to $g^d_L=-1/2+\sin^2{\theta}_w$ and $g^d_R=\sin^2{\theta}_w$
and  the  $g_p^f$ for the neutrino ($\nu_\alpha$ with $\alpha=e,\mu,\tau$)  corresponds to $g^d_L=1/2$ and $g^d_R=0$.
$\theta_w$ is the Weinberg mixing angle~\citep{1976NuPhA.274..368D,1988NuPhB.307..924N}, with a typical value of
$\sin^2{\theta_w}\approx 0.23$~\citep{2014ChPhC..38i0001O}.

\smallskip
The evolution of a generic neutrino state $\nu_\alpha \equiv ( \nu_e\; \nu_\mu\; \nu_\tau)^T$ is described by a 
Schr\"odinger-like equation~\citep{2008PhR...460....1G}, that expresses the evolution of the neutrino 
between the flavour states~\citep{2008PhR...460....1G}  with the distance $r$, 
from neutrinos that are produced in the Sun's core until their arrival  to the Earth's  neutrino detectors.
The equation reads
\begin{eqnarray}
i\frac{d\nu_\alpha}{dr}=\Hami \nu_\alpha = \left( \Hami_{v} +\Hami_{m} \right)\nu_\alpha,
\end{eqnarray}
where $\Hami$ is the total Hamiltonian,
$\Hami_v$ and $\Hami_m$ are the Hamiltonian components expressions for 
vacuum and in matter flavour variations, such that $\Hami_v \equiv M_\nu^{\dagger}M_\nu/2p_\nu $
where $M_\nu$ is the mass matrix of neutrinos (the term proportional to the neutrino momentum $p_\nu$
is omitted here), and  $\Hami_m $  is the Hamiltonian (a diagonal matrix of effective potentials) 
which depends on the properties of the solar plasma, i.e., the density and composition 
of the matter, such that $\Hami_m = diag (V_e,V_\mu,V_\tau) $.  

\smallskip
The flavour evolution is described in terms of the 
instantaneous eigenstates of the Hamiltonian in matter 
$\nu_m\equiv (\nu_{1m},\nu_{2m},\nu_{3m})^T$.  These eigenstates are related to the flavour states 
by the mixing matrix in matter, $U_m$: $\nu_\alpha=U^m\nu_m$.

\subsection{The effective matter potential}
As neutrinos propagate in the Sun's interior, they will oscillate between the three flavour states 
$\nu_e$, $\nu_\mu$ and $\nu_\tau$  due to vacuum oscillations, however in the highly dense medium which is the Sun's interior,
contrary to their propagation in vacuum, the scattering of neutrinos with other 
elementary particles, like electrons, will enhance their oscillation between flavour states.  
Indeed, neutrinos propagating in a dense medium like the Sun (or Earth) have their 
flavour between states affected by the coherent forward scattering, i.e., 
coherent interactions of the neutrinos with the medium background~\citep{2008PhR...460....1G}.
The interaction of neutrinos with the medium proceeds through coherent forward elastic Charged-Current ($cc$) 
and Neutral-Current ($nc$) scatterings, which as usual are represented by the  effective potentials
$V^{cc}_{\alpha}$ and $ V^{nc}_{\alpha}$ for each of the three type of neutrinos. 

\smallskip 
Therefore, at low energies, the potentials can be evaluated by taking the average 
of the effective four-fermion Hamiltonian due to exchange 
of $W$ and $Z$ bosons over the state describing the background medium.
Accordingly, for a non-relativistic un-polarized medium, 
for the effective potential of $\nu_e$, $\nu_\mu$ and $\nu_\tau$ neutrinos, 
one obtains
\begin{eqnarray}
V_\alpha= V^{cc}_{\alpha}+ V^{nc}_{\alpha},
\label{eq:Valpha1}
\end{eqnarray}
where $\alpha=e,\mu,\tau$.

\smallskip
Let us consider that the solar internal medium is mainly composed of electrons, 
up-quarks and down-quarks as in protons and neutrons with
the corresponding $n_e(r)$, $n_u(r)$ and $n_d(r)$ local number densities.
The contribution to ${\cal H}_m$ due to the $cc$ scattering of electron neutrinos 
$\nu_e$ (produced in the Sun's core) propagating in a homogeneous and isotropic gas of 
unpolarized electrons (like the electron plasma found in the Sun's interior) is given by 
\begin{eqnarray}
V^{cc}_e=\sqrt{2}G_F n_e (r)
\end{eqnarray}
where $G_f$ is the Fermi constant. 
 For $\nu_\mu$ and $\nu_\tau$, the potential due to its $cc$ interactions is zero for most of the solar interior 
since neither $\mu$'s nor $\tau$'s are present, therefore,
\begin{eqnarray}
V^{cc}_\mu=V^{cc}_\tau =0
\end{eqnarray} 

Generically, for any active neutrino, the $V^{cc}_\alpha$ reads
\begin{eqnarray}
V^{cc}_\alpha=\delta_{\alpha e} \sqrt{2}G_F n_e (r) 
\label{eq:Valphacc1}
\end{eqnarray}   

\smallskip
Analogously, one determines $V^{nc}_{\alpha}$ for any neutrino due to $nc$ interactions.
Since $nc$ interactions are flavour independent, these contributions are the same for neutrinos of all three flavours.
The neutral-current ($nc$) potential reads  
\begin{eqnarray}
V^{nc}_\alpha=\sum_{f} \sqrt{2} G_F g_v^f n_f (r)
\label{eq:Valphanc1}
\end{eqnarray}
where $\alpha=e,\mu,\tau$  and $f=e,u,d$. 
$n_f (r)$ is number density of fermions, electrons, up-quarks $(u)$ and down-quarks ($d$) as in protons ($uud$) and neutrons ($udd$). The factors $g_v^f$ are the axial coupling to fermions
($g_v^e=-1/2+2\sin^2{\theta_w}$, $g_v^u=1/2-4/3\sin^2{\theta_w}$ and $g_v^d=-1/2+2/3\sin^2{\theta_w}$,
see for example~\citet{2007fnpa.book.....G}).
%
Therefore the effective potential~\citep{2008PhR...460....1G} for any active neutrino  due to the neutral-current $V^{nc}_\alpha$
reads  
\begin{eqnarray}
V^{nc}_\alpha=\sqrt{2} G_F\left[g_v^e n_e(r) + g_v^u n_u(r) + g_v^d n_d(r)\right].
\label{eq:Valphanc2}
\end{eqnarray}
where $n_u(r)$ and $n_d(r)$ are the analogue of $n_e(r)$, i.e., 
the number density of up-quarks and down-quarks in the Sun's interior. 

\smallskip
Using equations (\ref{eq:Valphacc1}) and (\ref{eq:Valphanc1}) in 
equation (\ref{eq:Valpha1}), the effective potential for   any active neutrino crossing the solar plasma
reads 
\begin{eqnarray}
V_\alpha=  \sqrt{2} G_F \left[ \delta_{\alpha e}n_e + g_v^e n_e + g_v^u n_u + g_v^d n_d\right].
\label{eq:Valpha2}
\end{eqnarray}

\smallskip
When neutrinos propagate through matter, the forward scattering of neutrinos off the background matter
will induce an index of refraction for neutrinos. This is the exact analogous to the index of
refraction of light travelling through matter. However, the neutrino index of refraction will depend on the neutrino flavour, as the background matter contains different amounts of scatters for the different neutrinos flavours.   
    
The  effective potentials $V_\alpha$ are due to the coherent interactions of active flavour neutrinos
with the medium through coherent forward elastic weak $cc$ and $nc$ scatterings. 

\smallskip
Inside the Sun, as local matter is composed of neutrons, protons, and electrons,
the effective potential $V_\alpha$ for the different neutrino species
(including $\nu_e$ neutrinos) has a quite distinct form which depends on
the local number densities $n_e(r)$, $n_u(r)$ and $n_d(r)$, 
quantities which depend on the chemical composition (its metalicity $Z$) of the
Sun's interior. 
Nevertheless, at first approximation, since electrical neutrality implies locally
an equal number density of protons ($uud$) and electrons, $V_\alpha$ takes a more simple form
(equation~\ref{eq:Valpha2}), as the $nc$ potential contribution
of protons and electrons cancel each other. Therefore,
only neutrons ($udd$) contribute to  $V^{nc}_\alpha$.
Hence the last two terms of equation (\ref{eq:Valphanc2}) can be expressed  
as $g_v^n n_n(r)$ to only take into account the quark contribution for neutrons.
In this expression $n_n(r)$ is the local density of neutrons and
the $g_v^n$ is the neutron coupling constant, it follows that $g_v^n=g_v^u+2g_v^d=-1/2$,
and equation (\ref{eq:Valphanc2}) reads  $V_\alpha^{nc}=-\sqrt{2}/{2} G_F n_n (r)$.
Now $V_\alpha$ 
(equation~\ref{eq:Valpha2}) inside the Sun yields 
\begin{eqnarray}
V_\alpha=\sqrt{2} G_F \left[ \delta_{\alpha e} n_e (r) -\frac{1}{2} n_n(r)\right].
\label{eq:Valpha3}
\end{eqnarray} 
where $\alpha=e,\mu,\tau$.

\smallskip
As we will discuss later, only effective potential differences
affect the propagation of neutrinos in matter~\citep{1989RvMP...61..937K}, accordingly,
one defines the potential difference between two neutrino flavours 
$\alpha$ and $\beta$ as
\begin{eqnarray}
V_{\alpha\beta}=V_\alpha-V_\beta,
\label{eq:Valphabeta}
\end{eqnarray}
where $\alpha,\beta=e,\mu,\tau$.

\medskip

The  Sun's interior is a normal medium composed of 
nuclei (protons and neutrons)  and electrons.
Since  the effective potential for muon and tau neutrinos,
$V_\alpha$ (with $\alpha =\mu,\tau $ or a combination thereof)  
is due to the neutral current scattering only (see equation~\ref{eq:Valpha3}), 
this leads to $V_{\mu\tau}=V_\mu-V_\tau=0$. However, as the effective potential 
for electron neutrinos depends on the neutral and charged current scatterings,
in this case 
\begin{eqnarray}
V_{e\alpha}=\sqrt{2}G_Fn_e(r),
\label{eq:Vealpha}
\end{eqnarray}
where $\alpha =\mu,\tau $ or a combination thereof.
Although for the Sun and Earth only charged current interactions with electrons 
are the only effective potential that contributes to the propagation of electron neutrinos,
there are other types of non-typical matter, like the one found in the core of supernovae and in the early Universe for which the effective potential difference $V_{\alpha\beta}$ has a much stronger dependence on the properties of the background plasma~\citep{1988NuPhB.307..924N,1989RvMP...61..937K}.        

\subsection{Neutrino oscillation data parameters}
\label{sb-nodp}

As shown in the previous section, the neutrino flavour oscillations model is described 
with the help of 6 mixing parameters all of which are determined from experimental 
data~\citep{2016PhRvD..93e3016H}. The quantities are the following ones: 
the difference of the squared neutrino masses $ \Delta m^2_{21}$, $ \Delta m^2_{31} $,
the mixing angles  $\sin^2{\theta_{12}},\sin^2{\theta_{13}}, \sin^2{\theta_{23}} $ 
and the  CP-violation phase $\delta_{CP}$.

The mass square differences and mixing angles are known  with a good accuracy~\citep{2015JHEP...09..200B,2014JHEP...11..052G}:  $\Delta m_{31}^2$ is obtained from the experiments  of atmospheric neutrinos and $\Delta m_{12}^2$  
is obtained from solar neutrino experiments. 

The mixing angles are not  uniformly well defined: $\theta_{12} $ is
 obtained from solar neutrino experiments with an excellent precision; 
 $\theta_{23} $ is obtained from the atmospheric neutrino experiments, 
 this is the mixing angle of the highest value; $ \theta_{13} $ has been firstly estimated from the 
 Chooz reactor \citep{1999PhLB..466..415A}, its value is very small and 
 was still  very uncertain \citep{2009arXiv0905.3549F}.  Nowadays with Daya Bay and Reno, the situation has largely improved \citep{2012PhRvD..86a3012F}.
 However,  present experiments cannot fix 
 the value of  the CP-violation phase \citep{2010arXiv1010.4131H}. 

An overall fit to the data obtained from the different  neutrino experiments: 
solar neutrino detectors, accelerators, atmospheric neutrino detectors 
and nuclear reactor experiments suggests that  the parameters of neutrino oscillations are the following ones 
\citep{2015JHEP...09..200B,2014JHEP...11..052G}:
$ \Delta m_{31}^2\sim 2.457 \pm 0.045\; 10^{-3} eV^2 $ or
($ \Delta m_{31}^2\sim -2.449 \pm 0.048\; 10^{-3} eV^2 $) , 
$ \Delta m_{21}^2 \sim 7.500 \pm 0.019\; 10^{-5} eV^2$, 
$ \sin^2{\theta_{12}}=0.304 \pm 0.013 $  ,$ \sin^2{\theta_{13}}=0.0218 \pm 0.001 $,  $\sin^2{\theta_{23}}=0.562 \pm 0.032 $
and $\delta_{CP}=2\pi/25\; n$ with $n=1,\cdots,25$.  
  

In the limiting case where the value of the  mass differences, $ \Delta m^2_{12}$ or $ \Delta m^2_{31}$ is large, 
or one of the angles of mixing ($ \theta_{12}, \theta_ {23}, \theta_{31} $) is small, 
the theory of three neutrino flavour oscillations reverts to an effective theory of two neutrino flavour oscillations
\citep{2010arXiv1010.4131H}. 
Balantekin and Yuksel have shown that the survival probability of solar neutrinos 
calculated in a model with two neutrino flavour oscillations or three neutrino flavour oscillations have very close values \citep{2003hep.ph....1072B}.

\subsection{The survival of electron neutrinos}

Mostly motivated by solar neutrino data,  
the focus of this work is the study of the propagation of electron neutrinos,
in particular to determine the survival probability  of electron neutrinos 
$P_{e}$ ($\equiv P (\nu_e \rightarrow \nu_e) $ ) arriving on  Earth
which have their flavour changed due to vacuum and solar matter oscillations.  
Luckily, in the Sun $P_{e}$  takes a particularly simple form,
since the evolution of neutrinos in matter is  adiabatic
and for that reason their contribution for $P_{e}$ 
can be cast in a similar  manner to the vacuum-oscillation expression.
Accordingly, the standard parametrization of the neutrino mixing matrix 
leads to the following survival probability for electron neutrinos, 
$P_{e}$ reads 
\begin{eqnarray}
P_{e}=c^2_{13}c^{m2}_{13}P_2^{ad}+s^2_{13}s^{m2}_{13}
\label{eq:Pnuesm}
\end{eqnarray} 
where $c_{ij}=\cos{\theta_{ij}}$, $s_{ij}=\sin{\theta_{ij}}$, and
$P^{ad}_{2}$ reads
\begin{eqnarray}
P^{ad}_{2}=\frac{1}{2}\left(1+\cos{(2\theta_{12})}\cos{(2\theta^m_{12})} \right).
\end{eqnarray} 

\smallskip
The matter angles,  $\theta^{m}_{12}$ and  $\theta^{m}_{13}$ which depends equally of the fundamental
parameters of neutrino flavour oscillation and the properties of solar plasma are determined as follows:     

\begin{itemize}
\item 
The mixing angle $\theta^{m}_{12}$ is determined by 
\begin{eqnarray}
\cos{(2\theta^m_{12})}=-\frac{V^\star_{12}}{\sqrt{V_{12}^{\star 2}+(A_{12}^{\star-1}\sin{(2\theta_{12})})^2 } }
\end{eqnarray} 
where the effective potential $V^\star_{12}(E,r)$ reads
\begin{eqnarray}
V^\star_{12}(E,r)=c^2_{13}-A^{\star-1}_{12}\cos{(2\theta_{12})}
\end{eqnarray} 
where $A^\star_{12}=A_\star/\Delta m^2_{12}$ and $A_\star=2EV_{\alpha\beta}$~\citep{2008PhR...460....1G}. 
The parameter $A_\star (E,r)$ contains the effect of matter on the electron neutrino propagation
as defined by $V_{\alpha\beta}$, given by equation~(\ref{eq:Vealpha}).
In the specific case of electron neutrinos, $A_\star(E,r)=2E\; \sqrt{2}G_F n_e(r) $
(with $V_{e\alpha}=\sqrt{2}G_Fn_e(r)$). 

\item
The mixing angle $\theta^{m}_{13}$, accordingly to~\citet{2005PhRvD..72e3011G},
is determined by 
\begin{eqnarray}
\sin^2{(\theta^m_{13})}\approx \sin^2{(\theta_{13})}\left[1+2\;A^\star_{13}\right] 
\end{eqnarray} 
with $A^\star_{13}=2E V^o_e/\Delta m^2_{31}$, where $V^o_e$ is the effective potential
at the electron neutrino production radius $r_o$, i.e., $V^o_e=\sqrt{2}G_F n_e (r_o)$.
The value of $r_o$ is different for the different neutrino sources of the pp chains and CNO cycle. 
\end{itemize}

\subsection{New Neutrino Physics}
\label{sec-NNP}

In the presence of physics beyond the standard model~\citep[e.g.,][]{2013mpp..book.....T},
the neutral current interactions that are flavour diagonal and universal
in the standard model can have a more general form. Hence, new interactions arise 
between neutrinos and matter, which conveniently one defines as non-standard interactions ($nsi$),
these new neutrino interactions with fermions are described by a new
effective lagrangian~\citep[e.g.,][]{2003JHEP...03..011D,2013JHEP...09..152G,2015PhLB..748..311F}. 
Accordingly, the  classical lagrangian (equation~\ref{eq:Lagrsi}) is generalized to take into account
these new types of interactions previously forbidden. The new lagrangian reads
\begin{eqnarray}
\Lagr_{nsi}=-2\sqrt{2}G_F
\epsilon_{\alpha\beta}^{fP} 
\left( \bar{\nu}_\alpha\gamma_\rho L\nu_\beta \right)
\left( \bar{f}\gamma^\rho P f \right)
\label{eq-Lagr_nsi}
\end{eqnarray}
where $ \epsilon_{\alpha\beta}^{fP} $ is the equivalent of  $g_p^f$ for the standard interactions (equation~\ref{eq:Lagrsi}), which corresponds to the 
parametrization of the strength of the non-standard interactions
between neutrinos of flavours $\alpha$ and $\beta$ and the $P$-handed component 
of the fermion $f$~\citep[e.g.,][]{2013mpp..book.....T}. Without loss of generality 
we consider only neutrino interactions with up- and down-quarks~\citep[e.g.,][]{2016EPJA...52...87M}.
In the latter Lagrangian, $ \epsilon_{\alpha\beta}^{fP} $  corresponds to two classes of non-standard terms: 
flavour preserving non-standard terms proportional to $\epsilon_{\alpha\alpha}^{fP}$
(known as non-universal interactions),  and  flavour changing terms proportional to
$\epsilon_{\alpha\alpha}^{fP}$ with $\alpha\ne\beta$.

\smallskip
Since the atoms and ions of the solar medium in which neutrinos propagate are non-relativistic, 
the vector part of the $nsi$ operator gives the dominant contribution for the interactions
of the neutrinos with the plasma of the Sun's interior, in which case the effective $nsi$ coupling 
can be described by the following combination~\citep{2013JHEP...09..152G}:
$\epsilon^f_{\alpha\beta}=\epsilon^{fL}_{\alpha\beta}+\epsilon^{fR}_{\alpha\beta}$.
These new kinds of neutrino interactions lead to a new effective potential
difference to describe the propagation of neutrinos in matter~\citep{2016EPJA...52...87M}. 
Accordingly, the effective potential difference
$V_{\alpha\beta}$ is written as a generalization of the $V_{\alpha\beta}$ obtained in the 
standard case: equations~\ref{eq:Valpha3} and~\ref{eq:Vealpha}.
Hence, $V_{\alpha\beta}$ reads
\begin{eqnarray}
V_{\alpha\beta}=V_e\delta_{\alpha e}\delta_{\beta e}+\sqrt{2}G_F\sum_{f} \epsilon^f_{\alpha\beta}n_f(r)
\end{eqnarray}
where $\epsilon^f_{\alpha\beta}$ is the strength of $nsi$ interaction of neutrinos with the medium.
A more detailed discussion about the relations between $\epsilon^f_{\alpha\beta}$ 
and $\epsilon_{\alpha\beta}^{fP}$ can be found in~\citet{2013JHEP...09..152G}. 
Usually, $\epsilon^f_{\alpha\beta}$ is considered as a free parameter to be adjusted to fit the solar observational data.

\smallskip
In this work, we study only the $nsi$ interactions of electron neutrinos ($\nu_e$) with the solar plasma.   
Accordingly, as is common practice, we chose to take into account only the $nsi$ coupling of electron neutrinos 
with the up-quarks and down-quarks of the solar plasma.  Among others, 
\citet{2004PhRvD..70k1301F}~have shown that the coupling of electron neutrinos with  
up-quarks is parametrized by a set of two independent parameters ($\epsilon^u_N,\epsilon^u_D$),
and similarly the coupling of electron neutrinos with down-quarks is parametrized by 
another set of two independent parameters ($\epsilon^d_N,\epsilon^d_D$).  
Each of these parameters corresponds to a linear combination of 
the original parameters $\epsilon_{\alpha\beta}^f$ which defines 
the strength of the non-standard neutrino interactions  with fermions
as defined in equation~\ref{eq-Lagr_nsi}. In the appendix~\ref{sec-Ap}
we show the relation of $\epsilon_{D}^f$ and $\epsilon_{N}^f$  
with the parameters $\epsilon_{\alpha\beta}^f$, for which $f$ is either $d$ or $u$
since in our study we are only concerned about the interaction with down- and up-quarks of the solar plasma.
A detailed account of the relevance of these quantities can be found in~\citet[e.g.,][]{2004PhRvD..70k1301F,2013JHEP...09..152G,2016EPJA...52...87M}.

\smallskip
As in the case of standard neutrino interactions, for these $nsi$ interactions
the oscillations of neutrino flavour are still adiabatic,  so the probability of electron neutrino 
survival is given by equation (\ref{eq:Pnuesm}). However, in this case the quantity $\cos{(2\theta_m)}$ 
has been redefined to take into account the new effective matter potential~\citep{2016EPJA...52...87M}, 
accordingly   
\begin{eqnarray}
\cos{(2\theta^m_{12})}\approx - \frac{V^\star_{nsi}}{\sqrt{V_{nsi}^{\star 2}+ 
(2r_f \epsilon^f_N +A_{12}^{\star-1} \sin{(2\theta_{12})})^2 } }
\label{eq:costhetanis}
\end{eqnarray} 
where $V_{nsi}^{\star}$ reads
\begin{eqnarray}
V^\star_{nsi}(E,r)=c^2_{13}-A_{12}^{\star-1} \cos{(2\theta_{12})}- 2r_f  \epsilon^f_D. 
\label{eq:Vstarnis}
\end{eqnarray} 
where $r_f(r)=n_f(r)/n_e(r)$. 
\begin{figure}[!t]
\centering 
\includegraphics[scale=0.45]{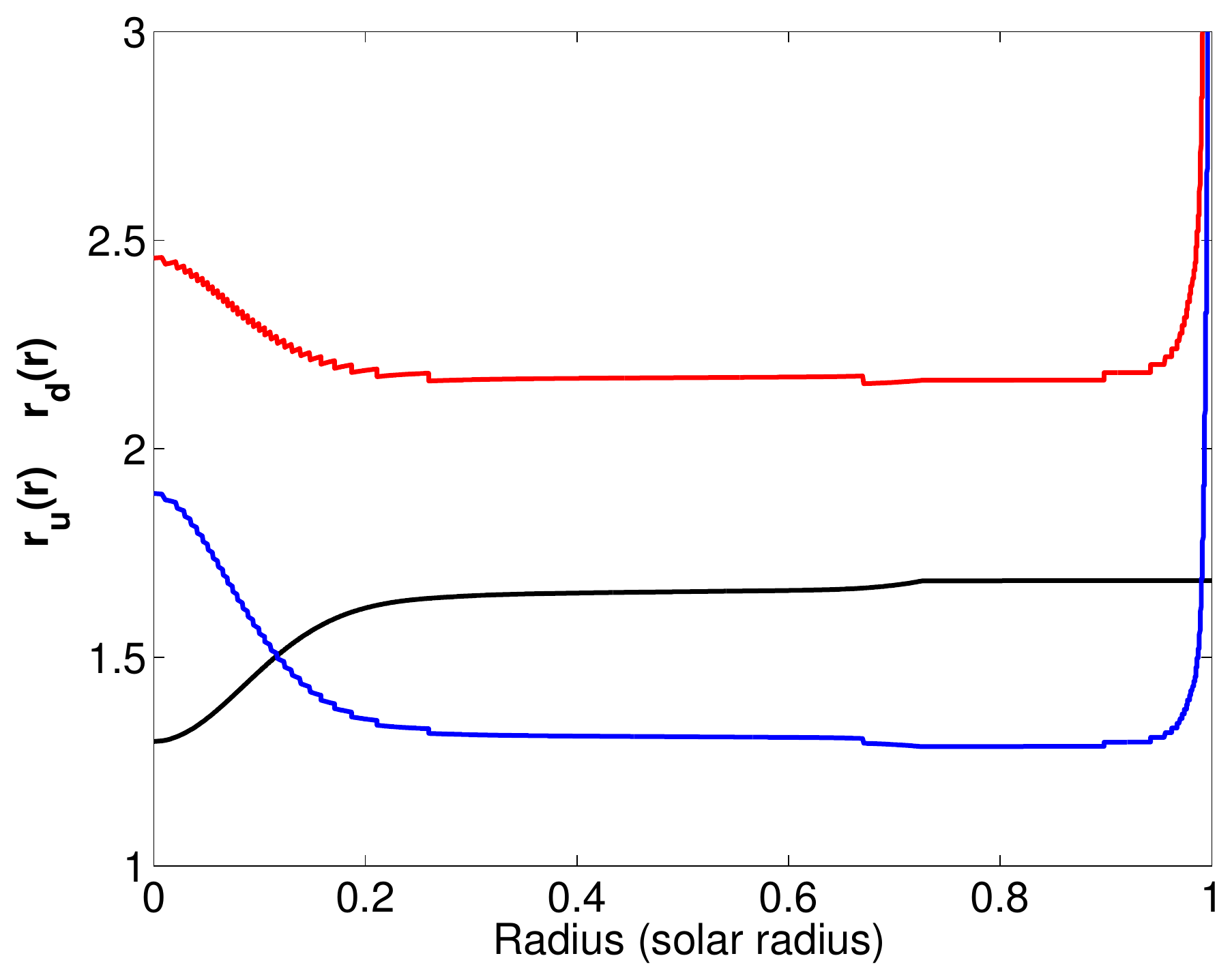}
\caption{Variation with the solar radius  of the ratios $r_u(r)=n_u(r)/n_e(r)$ (red curve)
and $r_d(r)=n_d(r)/n_e(r)$ (blue curve), and relative variation of 
up and down quarks $n_u(r)/n_d(r)$  (black curve).}
\label{figure:3}
\end{figure}
Figure~\ref{figure:3} shows the variation of the ratios $r_u(r)$ and $r_d(r)$ inside
the star. The $r_d(r)$ is smaller than $r_u(r)$ because the star's composition is dominated by free protons 
(ionized hydrogen). As such, for each down-quark there are two up-quarks.
\begin{figure}[!t]
\centering 
\includegraphics[scale=0.40]{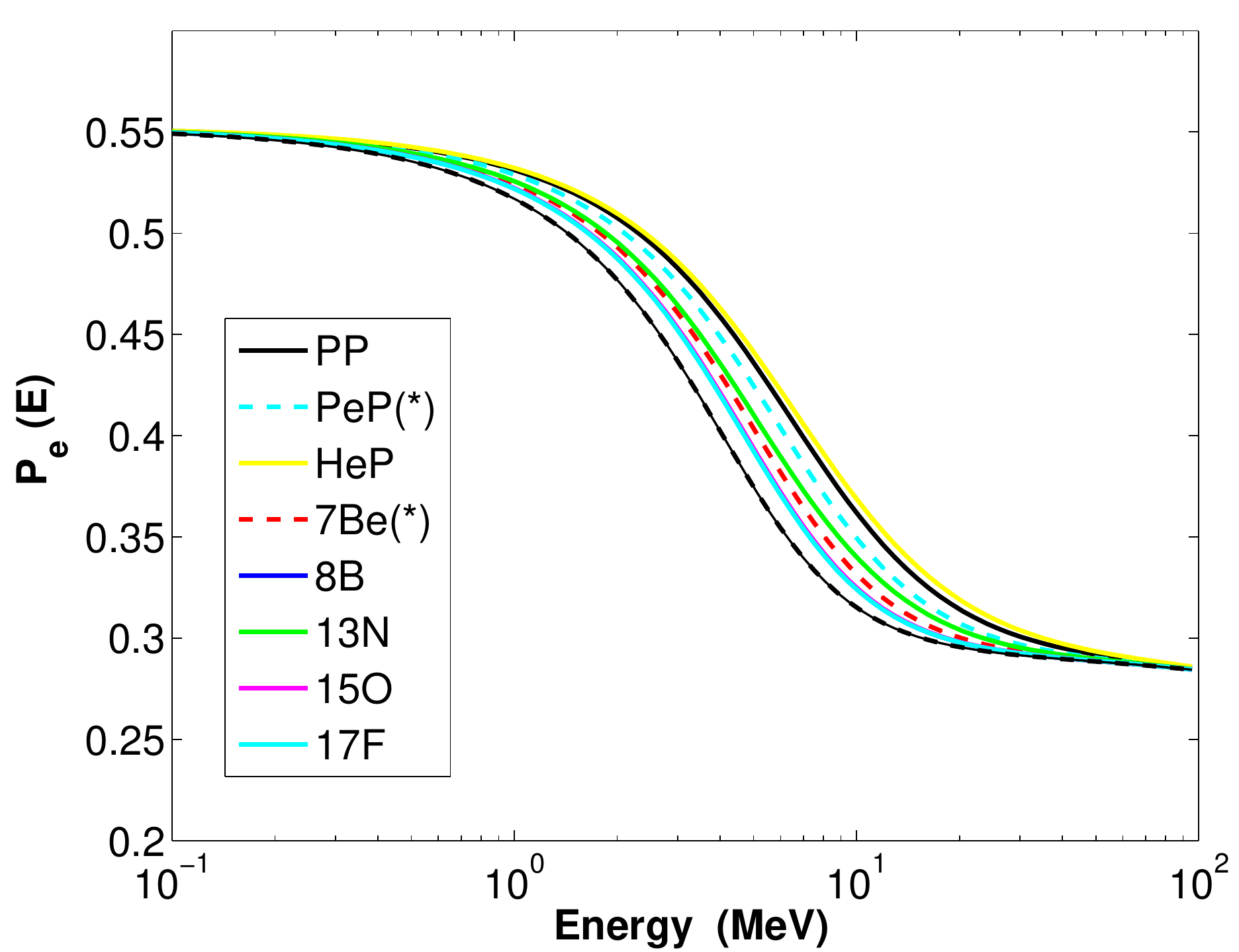}
\includegraphics[scale=0.40]{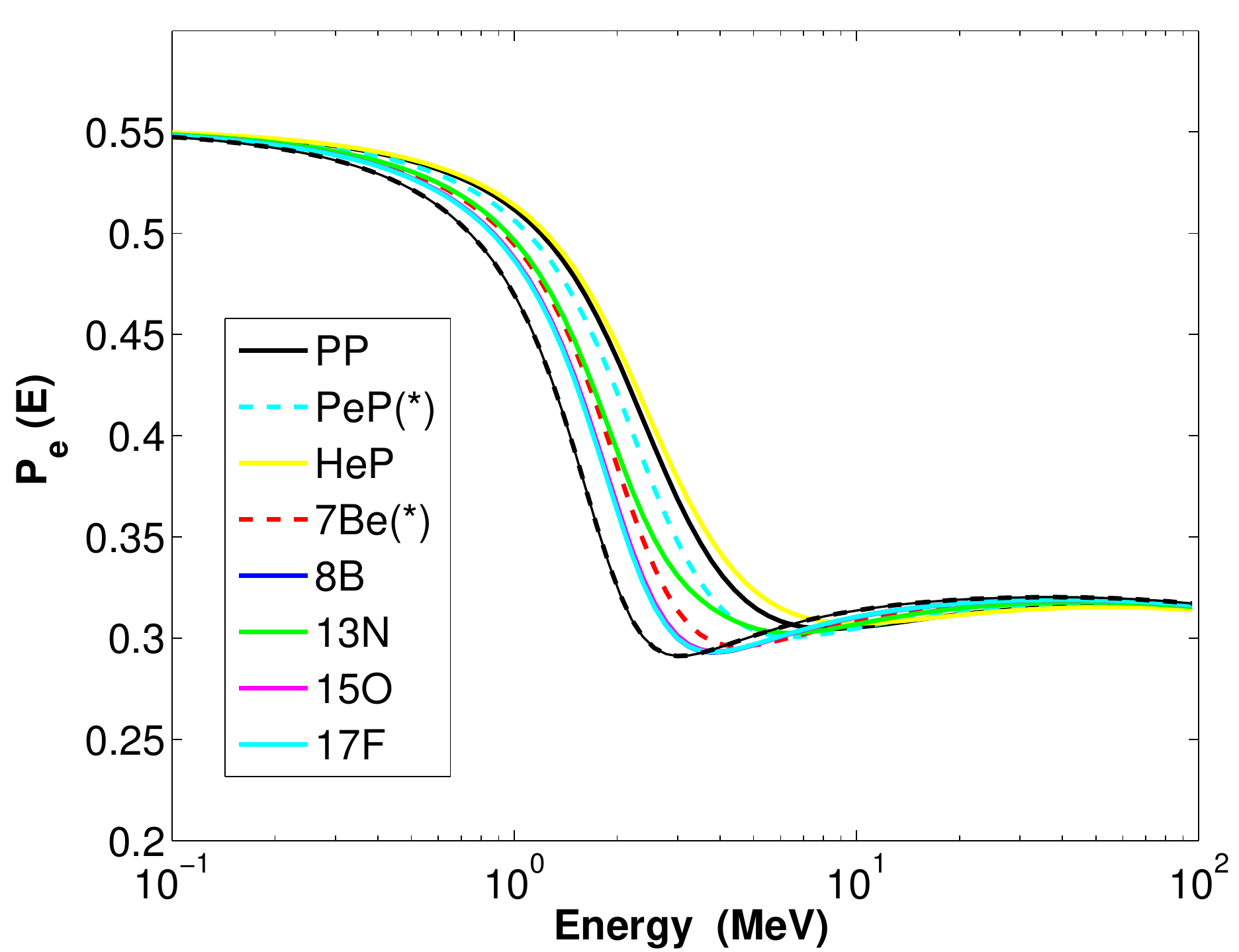}
\includegraphics[scale=0.40]{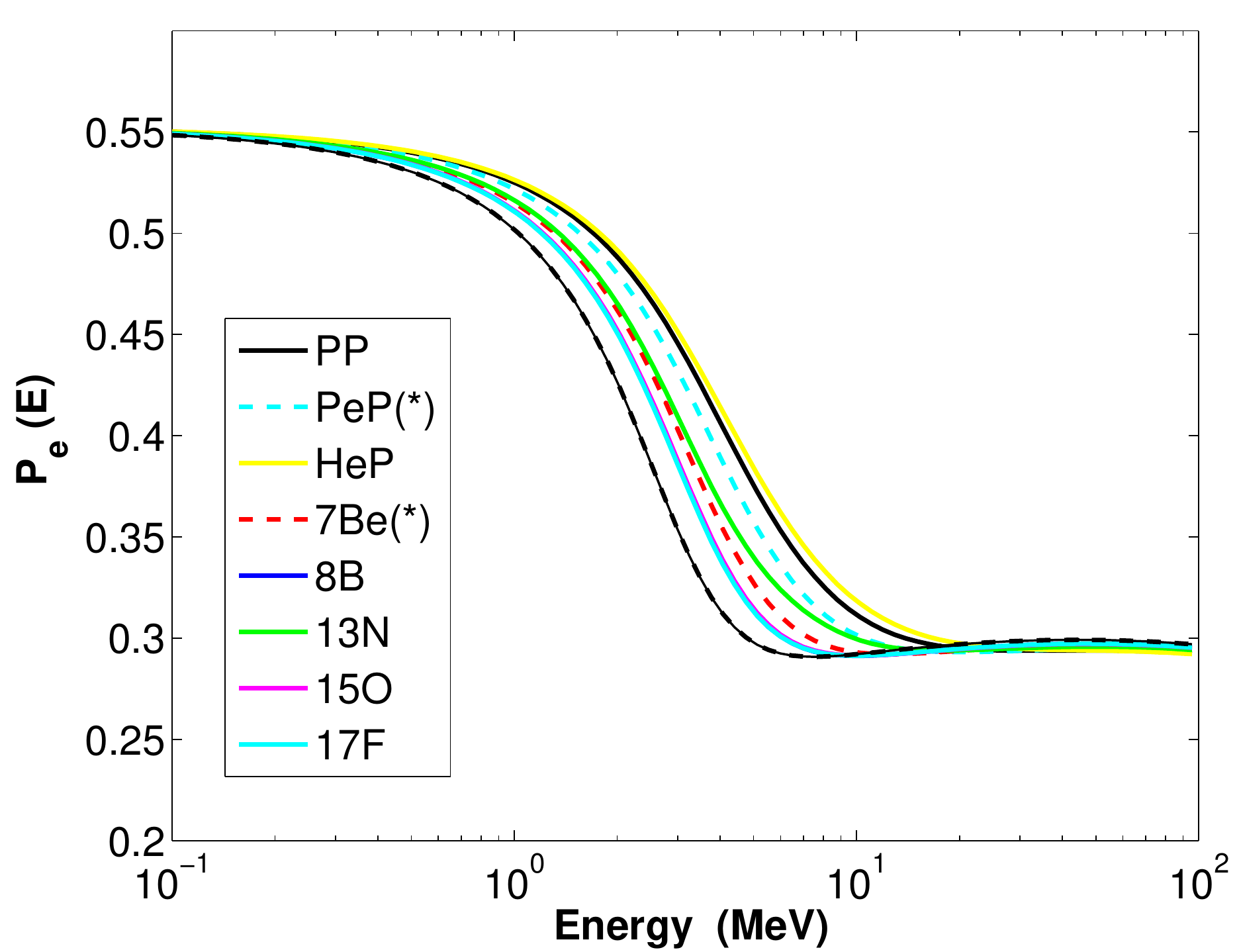}
\caption{The survival probability of electron-neutrinos: the $P_{e}$ curves correspond to neutrinos 
produced in the nuclear reactions located at different solar radius. The three panels correspond to the following
neutrino models of interaction: $si-$interaction with electrons ({\it top panel}),  
$nsi-$interaction with up-quarks with the coupling constants, $\epsilon^u_N=-0.30$ and $\epsilon^u_D=-0.22$
({\it middle panel}), and $nsi-$interaction with down-quarks with coupling constant, 
$\epsilon^d_N=-0.16$ and $\epsilon^d_D=-0.12$ ({\it bottom panel}). 
The reference dotted-black curve defines the survival probability of electron-neutrinos in the centre of the Sun 
for which the $si-$ or $nsi-$MSW flavour oscillation mechanism is maximum. 
The other coloured curves follow the same colour scheme shown in Figure~\ref{figure:1}. }
\label{figure:4}
\end{figure}  

\begin{figure}[!t]
\centering 
\includegraphics[scale=0.45]{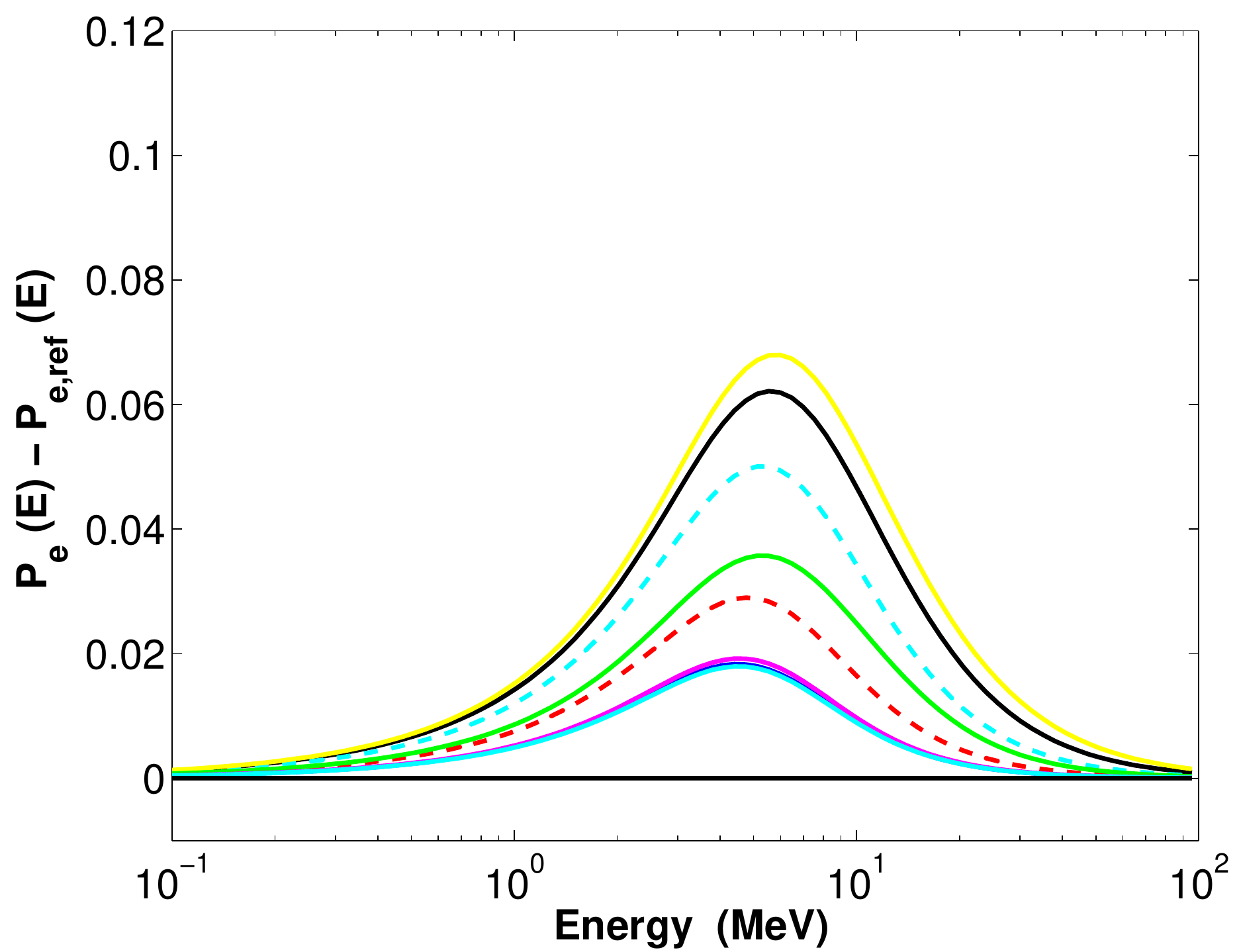}
\includegraphics[scale=0.45]{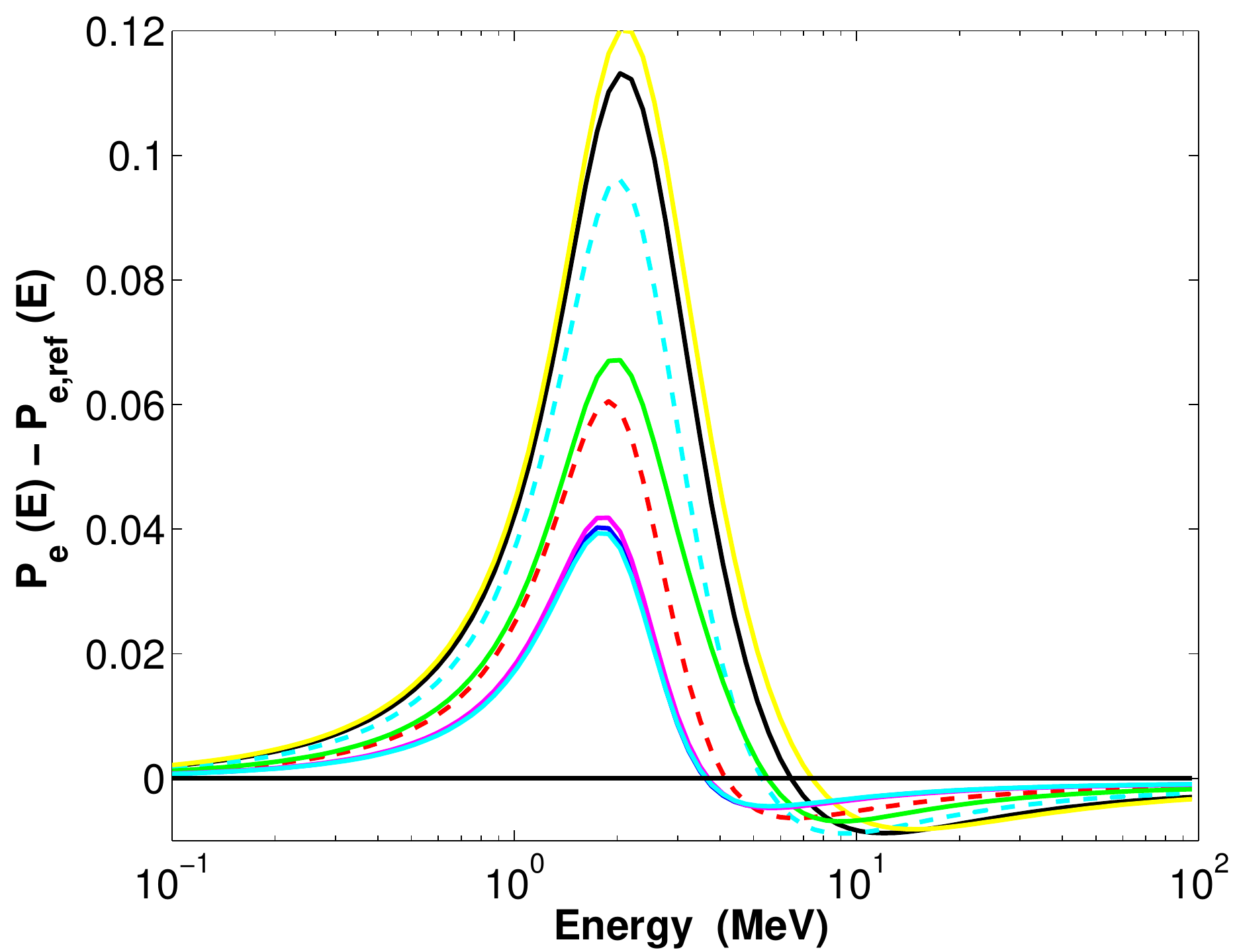}
\includegraphics[scale=0.45]{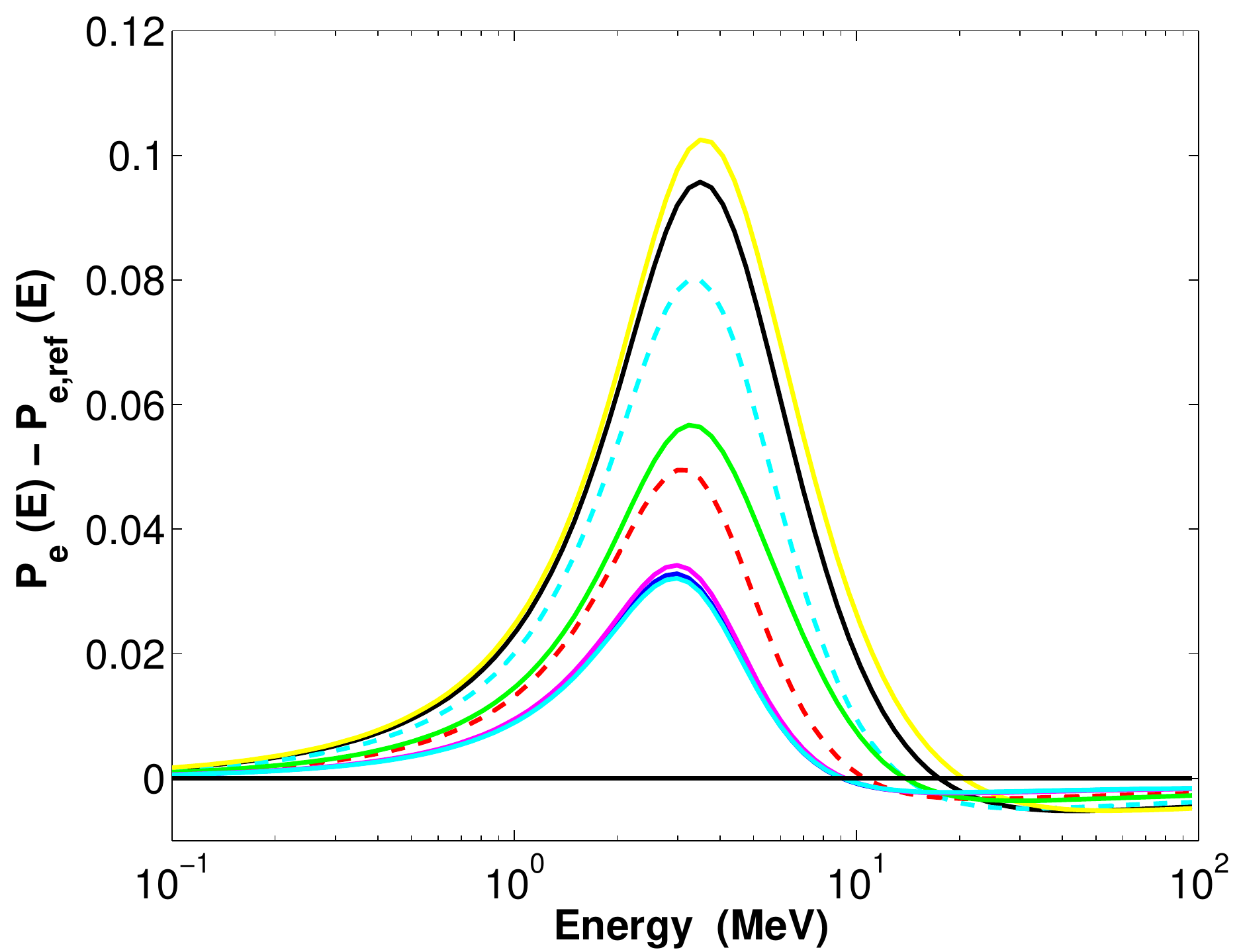}
\caption{The survival probability of electron-neutrinos in function of the neutrino energy 
for the different regions of emission. The different panels show the
difference between the survival probability of the different electron neutrino sources
and the reference curve (dotted-black curve in Figure~\ref{figure:4}).
The coloured curves follow the same colour scheme shown in Figures~\ref{figure:1} and~\ref{figure:4}.}
\label{figure:5}
\end{figure}

\subsection{The electron neutrino probability survival}

The neutrino emission reactions of  the pp chains and the CNO cycle are produced at high temperatures
in distinct layers in the Sun's core. Similarly, the neutrino flavour oscillations occur in the same 
regions. The average survival probability of electron neutrinos in each nuclear reaction region is given by
\begin{eqnarray} 
\langle P_{e} (E)\rangle_j = 
N_j^{-1} \int_0^{R_\odot} P_{e} (E,r)\phi_j (r) 4\pi \rho(r) r^2 dr 
\label{eq:Pnuej}
\end{eqnarray}  
where $ N_j $ is a normalization constant given by  $ N_j= \int_0^{R_\odot}\phi_j (r) 4 \pi \rho (r) r^ 2  \;dr $ 
and $ \phi_j (r) $  is the electron neutrino emission function for the $j$ nuclear reaction. $j$ corresponds 
to the following electron neutrino nuclear reactions: $PP$, $PeP$, $^8B$, $^7Be$, $^{13}N$, $^{15}O$ and $^{17}F$.
$\phi_j(r)$ defines the location where neutrinos
are produced in each nuclear reaction $j$  for which the production is maximum in the layer 
of radius $\rm r_j$ (Cf. figure~\ref{figure:1}).
The neutrino fluxes produced by the different nuclear reactions are sensitive 
to the local values of the temperature, molecular weight,  density and electronic density. 
In this study, we consider that all neutrinos produced in the solar nuclear reactions are of 
electron flavour as predicted by standard nuclear physics, therefore, the local density of quarks only affects the $\langle P_{e} (E)\rangle_j $ by modifying  the flavour of electron neutrinos by a new $nsi$ interaction like the generalized MSW mechanism.      
 
\smallskip
The survival probability of electron neutrinos $\langle P_{e} (E)\rangle_j$  given by equation (\ref{eq:Pnuej})
is computed using equations (\ref{eq:Pnuesm}), (\ref{eq:costhetanis}) and (\ref{eq:Vstarnis}). 
Figures~\ref{figure:4} and~\ref{figure:5} show $\langle P_{e} (E)\rangle_j$ for the different solar neutrino sources,
either in the standard MSW or a generalized MSW. The different neutrino interaction models are described 
by a specific set of parameters: ($\epsilon^u_N,\epsilon^u_D,\epsilon^d_N,\epsilon^d_D$). 
Figures~\ref{figure:4} and~\ref{figure:5} top panels show $\langle P_{e} (E)\rangle_j$ for
the standard MSW mechanism in which case all the parameters mentioned above are equal to zero. 
The other panels of Figures~\ref{figure:4} and~\ref{figure:5} correspond to a generalized MSW mechanism 
for which the parameters ($\epsilon^u_N,\epsilon^u_D,\epsilon^d_N,\epsilon^d_D$) can have values different of zero. 
\citet{2016EPJA...52...87M} among others have shown that only a relatively small ensemble
of parameter combinations ($\epsilon^u_N,\epsilon^u_D,\epsilon^d_N,\epsilon^d_D$)
can be accommodated with  the current set of neutrino flux observations.
In this study, for convenience, we choose to focus on neutrino interactions
for which electron neutrinos couples either with up-quarks (for which $\epsilon^d_N=\epsilon^d_D=0$)  
or with down-quarks (for which $\epsilon^u_N=\epsilon^u_D=0$).
Specifically, we chose two fiducial sets of values ($\epsilon^f_N,\epsilon^f_D$, $f=u,d$))
of  the ensemble of parameters that  fits simultaneously the solar and KamLAND neutrino 
data sets with good accuracy. Figure~\ref{figure:4} shows the survival probability of electron-neutrinos
in the case of the $nsi-$interaction for the parameters sets: $(\epsilon^u_N=-0.30,\epsilon^u_D=-0.22)$
and   $(\epsilon^d_N=-0.16,\epsilon^d_D=-0.12)$. As discussed by~\citet{2016EPJA...52...87M} 
these values correspond to the two parameters  that best fit simultaneously the current solar and KamLAND neutrino data sets.
Figures~\ref{figure:4} and~\ref{figure:5} middle panels show $\langle P_{e} (E)\rangle_j$ for
a neutrino up-quark interaction model, and Figures~\ref{figure:4} and~\ref{figure:5} bottom panels 
show $\langle P_{e} (E)\rangle_j$ for a neutrino down-quark interaction model.
     
\smallskip
All the different neutrino interaction models have several common features. 
In general the $\langle P_{e} (E)\rangle_j$  are very similar for low- and high-energy neutrinos.
It is only for neutrinos with intermediate energy that it is possible to distinguish between 
the different models (Cf. figure~\ref{figure:4}). For neutrinos in this energy interval  
it is possible to distinguish two effects: 
one relates with the location of the different neutrino sources, and a second effect relates with 
the parameter values ($\epsilon^u_N,\epsilon^u_D,\epsilon^d_N,\epsilon^d_D$) of the neutrino interaction model.
In the former effect, the $\langle P_{e} (E)\rangle_j$ differentiation results from the fact that $\phi_j(r)$ 
are located at different solar radius, as shown in figure~\ref{figure:1}. 
Such effect arises equally in $si-$ and $nsi-$ neutrino interaction models.
The second effect occurs only for $nsi-$ neutrino interaction models, 
and it is related with the radial distribution of up- and down-quarks (cf. Figure~\ref{figure:2}).

\smallskip    
This latter effect is shown in figures~\ref{figure:4} and~\ref{figure:5} for the  
two fiducial models adopted in this study: a pure neutrino up-quark model ($\epsilon^d_N=\epsilon^d_D=0$)  
and a pure neutrino down-quark model ($\epsilon^u_N=\epsilon^u_D=0$). 
Accordingly, for neutrinos  with an intermediate energy,
the $\langle P_{e} (E)\rangle_j$  corresponding to the neutrino up-quark interaction model 
has an impact of larger amplitude than the neutrino down-quark interaction model 
(compare middle and bottom panels in figure~\ref{figure:5}). For instance, this effect 
is very significant for the $\langle P_{e} (E)\rangle_{pp}$. Nevertheless, these preliminary results 
should be interpreted with caution, since each neutrino source only produces neutrinos within a limited range of energy, as such the $nsi-$effect of $\langle P_{e} (E)\rangle_{j}$ shown in figure~\ref{figure:5} can be significantly reduced in the final neutrino spectrum of some solar neutrino sources. 
Indeed, we remind that the neutrinos emitted by $ \phi_j (r) $ are limited to a specific energy range for each $j-$ nuclear reaction. As such only an energy portion of $\langle P_{e} (E)\rangle_j$ affects the final emitted neutrino spectrum. This point will be discussed in more detail in the next section.  

\section{The solar electron neutrino spectra}
\label{sec-SEMNS}
The solar energy spectrum of electron neutrinos from any specific nuclear reaction
is known to be essentially independent of solar parameters, that is, the
energy spectrum created by a specific nuclear reaction is the same independently 
of whether neutrinos are produced in an Earth laboratory or in the core of the Sun. 
Therefore, the neutrino energy spectrum of the different nuclear reactions, 
can be assumed to be equivalent to its Earth laboratory counterpart.  
A typical example of such spectra is the $^8$B neutrino energy spectrum emitted by 
the $^8$B nuclear reaction of the pp chains in the Sun's core.  
This solar neutrino spectrum has been shown to be equivalent to  several experimental determinations of the 
$^8$B neutrino spectrum~\citep[e.g.][]{2000PhRvL..85.2909O,2006PhRvC..73b5503W}. 
\citet{1986PhRvC..33.2121B,1987PhRvC..36..298N}, among others, have shown that 
the $^8$B neutrino spectrum emitted in the Sun's core is  equal to the spectrum measured in the laboratory,
as the surrounding solar plasma does not affect this type of nuclear reaction.
The $^8$B neutrino spectrum measured in the laboratory agrees remarkably well with its theoretical 
prediction for neutrinos with an energy below 12 MeV,  
a small difference appearing only for high energy neutrinos.
The experimental $^8$B neutrino spectrum deduced from four laboratory experiments 
shows a difference with the theoretical prediction   
at most of 1\%~\citep{2003PhRvL..91y2501W,2006PhRvC..73b5503W,2012PhRvL.108p2502R,2006PhRvC..73e5802B,2011PhRvC..83f5802K,2011PhRvC..83f5802K,2012PhRvL.108p2502R}.
Accordingly, we will consider that the electron neutrino energy spectrum of a solar nuclear reaction at the 
specific location where these neutrinos are created  is identical to the equivalent neutrino spectrum measured in the laboratory.  

\begin{figure*}[t!]
\centering
\mbox{\subfigure[$\nu_e(PP)$]{\includegraphics[width=2.4in]{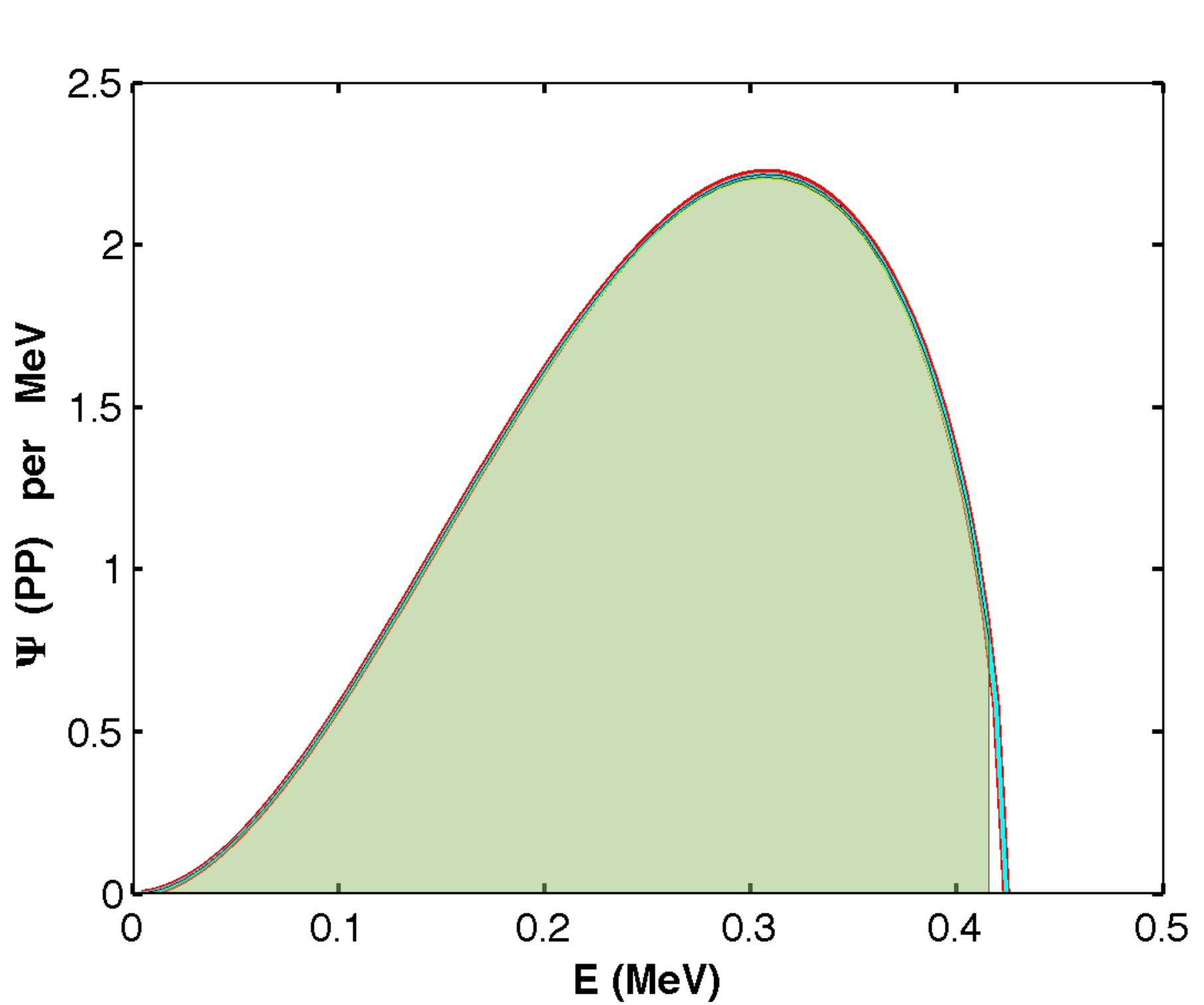}}\quad\hspace{-0.5cm}
\subfigure[$\nu_e(HeP)$]{\includegraphics[width=2.4in]{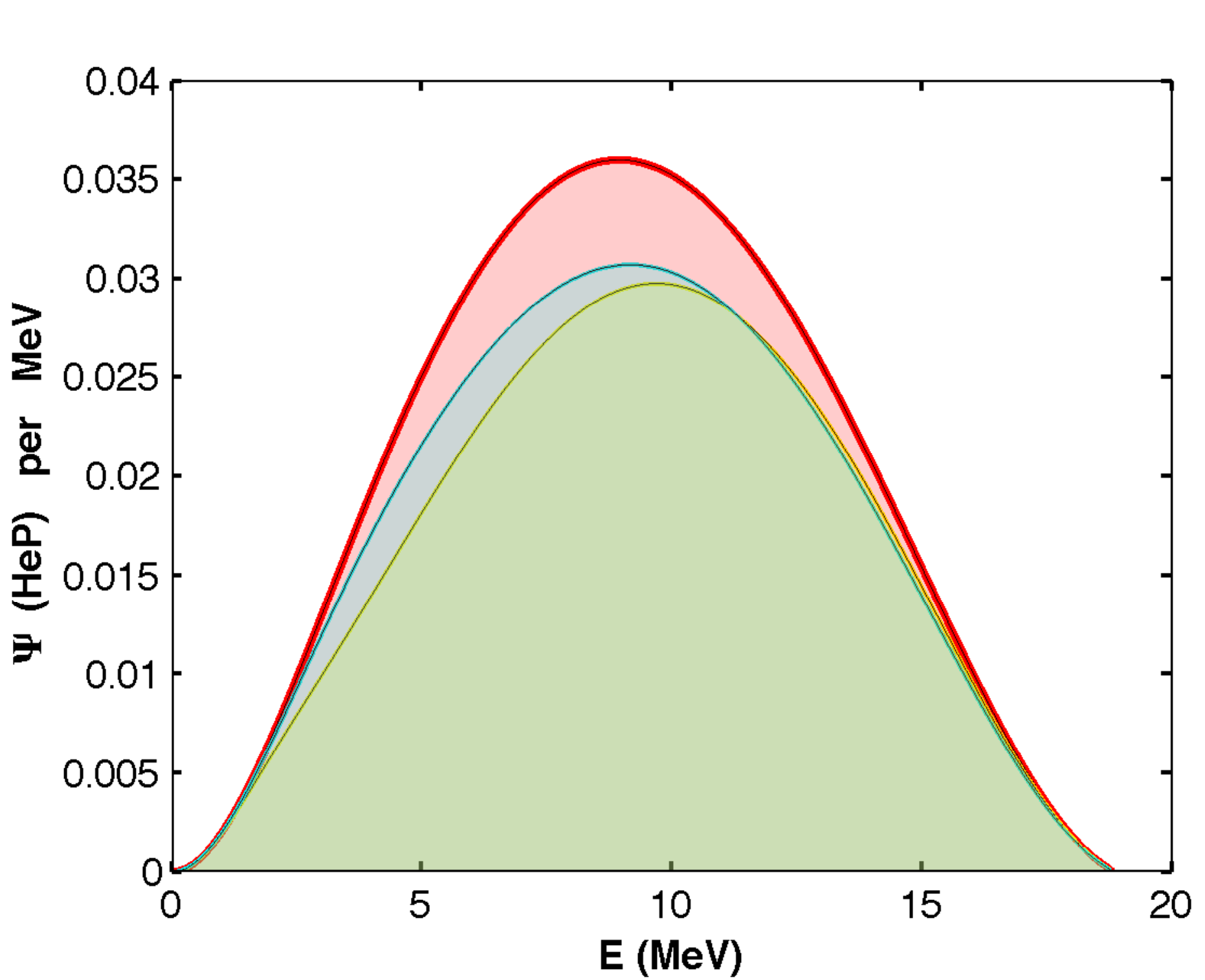}}\quad\hspace{-0.5cm}
\subfigure[$\nu_e(^8B)$]{\includegraphics[width=2.4in]{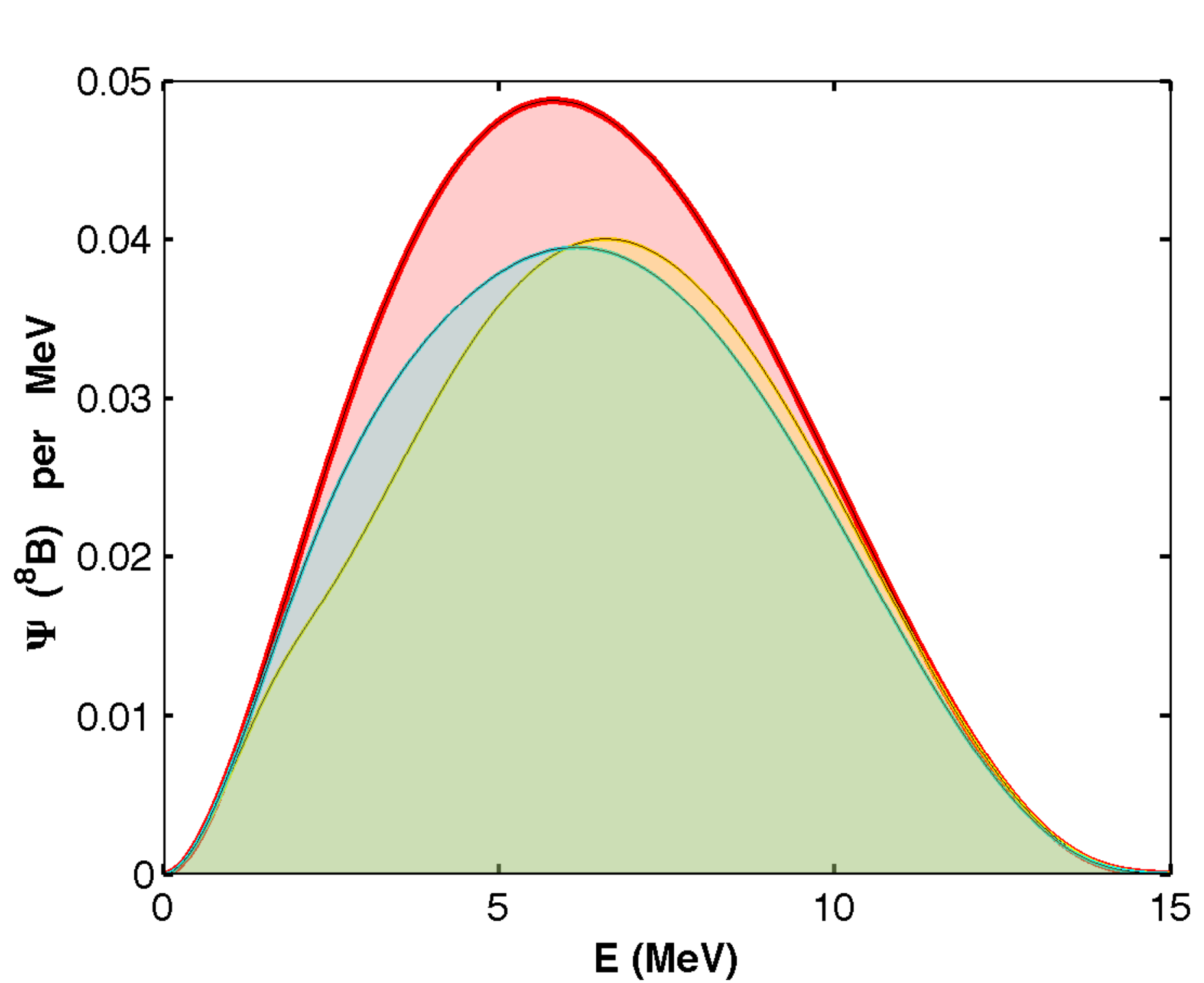}}}
\mbox{\subfigure[$\nu_e(^{13}N)$]{\includegraphics[width=2.4in]{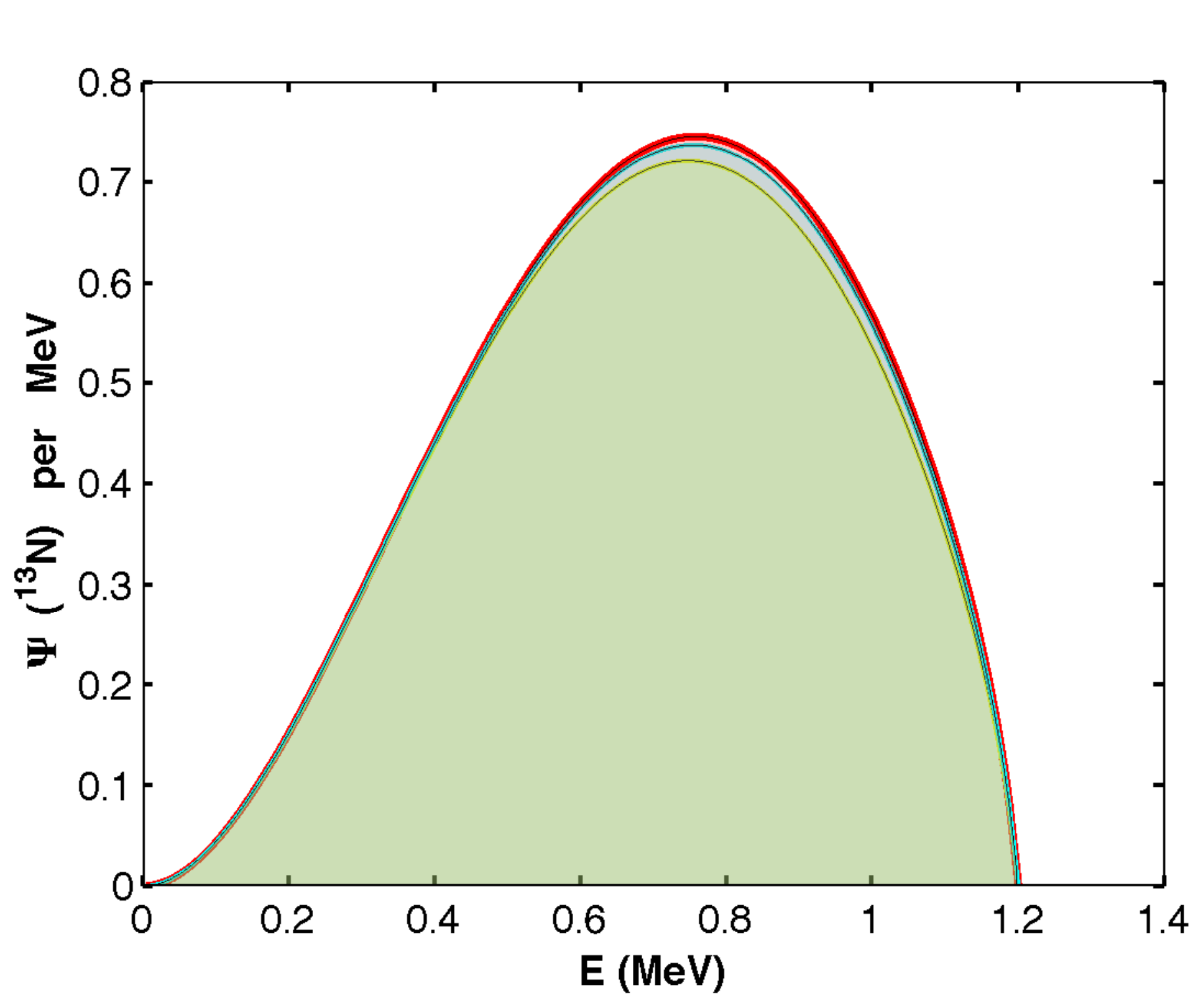}}\quad\hspace{-0.5cm}
\subfigure[$\nu_e(^{15}O)$]{\includegraphics[width=2.4in]{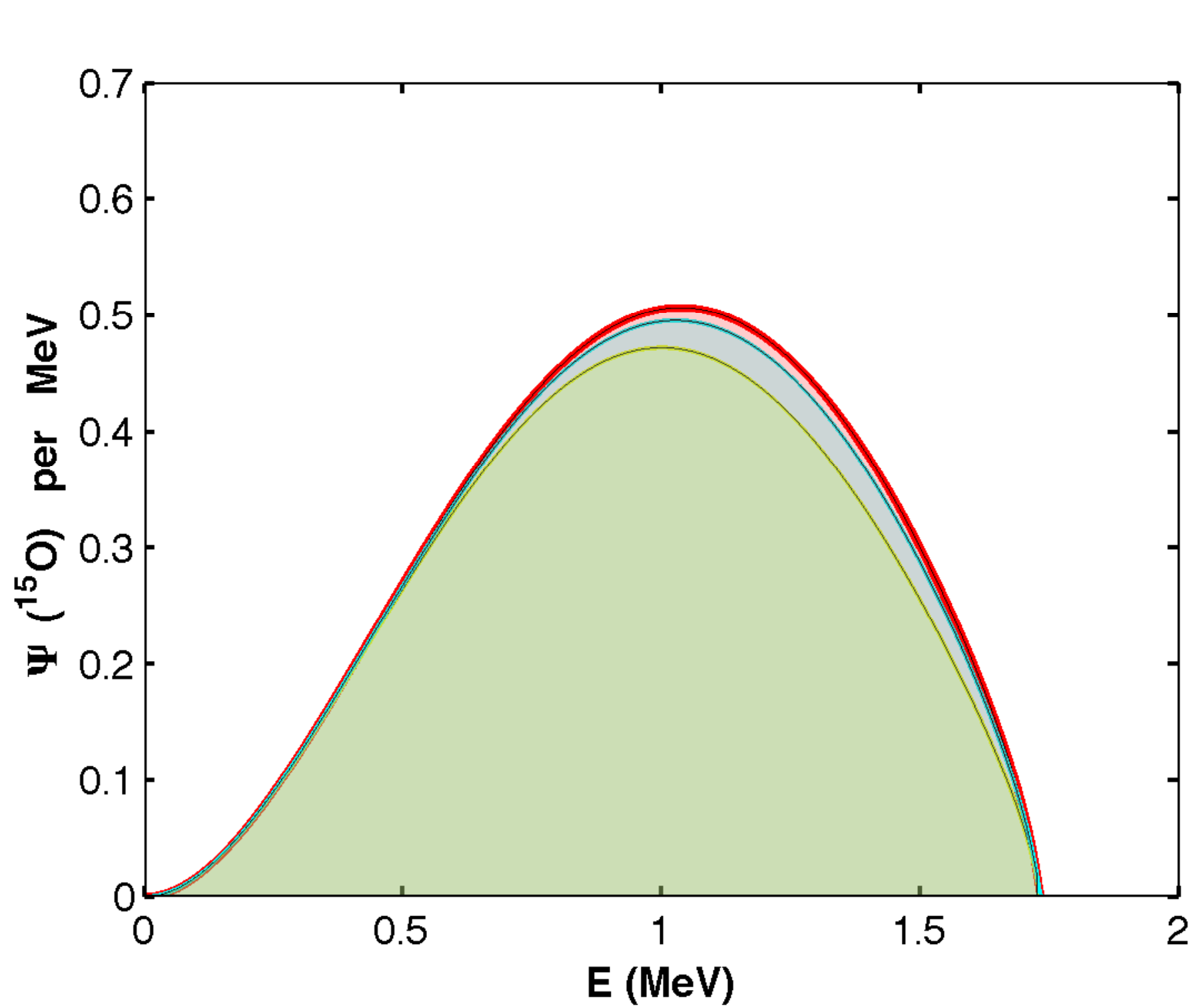}}\quad\hspace{-0.5cm}
\subfigure[$\nu_e(^{17}F)$]{\includegraphics[width=2.4in]{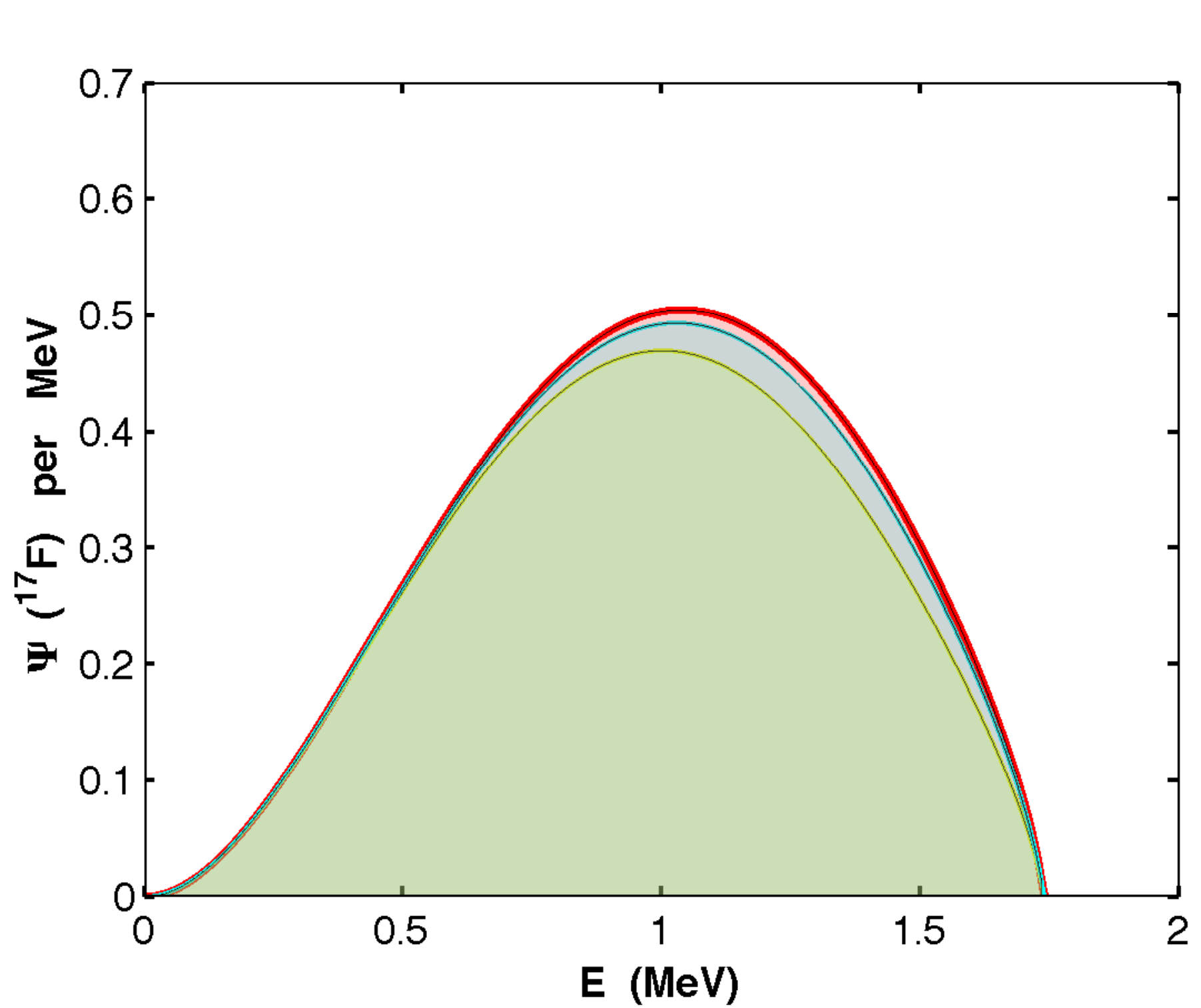}}}
\caption{$pp$-chain and $cno$-cycle energy neutrino spectra:
$\Psi_{e,j}^{\odot} (E)$ is the electron solar neutrino spectrum for the standard MSW effect (red area);
the other colour curves correspond to $\Psi^{e,j}_{\odot} (E)$, for which the
generalized MSW effect is taken into consideration:
$nsi-$neutrino interaction with up-quark (blue area), and
$nsi-$neutrino interaction with down-quarks (green area).
$\Psi_{e,j}^{\odot} (E)$ is defined as the probability per MeV of an electron-neutrino with an energy $E$. 
In the calculation of these neutrino spectra we used an up-to-date standard solar model.} 
\label{figure:6}
\end{figure*}

\smallskip
All neutrinos produced in the Sun's nuclear reactions are of electron neutrino type. 
It is only during the propagation phase that these neutrinos vary their flavour between electron,
$\tau$  and $\mu$. Suitably, we define the original energy spectrum of electron neutrinos by $\Psi^s_{e}(E)$  
and the end energy spectrum of neutrinos after the neutrino flavour oscillations by  $\Psi^\odot_{e}(E)$. The first spectrum, which is identical to the neutrino spectrum obtained in an Earth's laboratory,  relates to the neutrinos produced in nuclear reactions (see figure~\ref{figure:1}). The latter spectrum corresponds to neutrinos that have their flavour modified by the  vacuum oscillations and the  generalized MSW oscillation mechanism.

Conveniently, the neutrino spectra $\Psi^s_{e}(E)$  and $\Psi^\odot_{e}(E)$  associated 
to each of the different solar nuclear reactions of the $pp$-chain and CNO-cycle are labelled 
by an unique subscript $j$ which can take one of these values: $PP$, $HeP$, $^8B$, $^{13}N$, $^{15}O$ and $^{17}F$. 
Hence, the two previous neutrino energy spectra have a simple relation, it reads 
\begin{eqnarray} 
\Psi^\odot_{e,j}(E)=\langle P_{e} (E)\rangle_j \Psi^s_{e,j}(E),
\label{eq-Psie}
\end{eqnarray}
where $\langle P_{e} (E)\rangle_j$ is the  survival probability of an electron neutrino of energy $E$.
Figure~\ref{figure:6} shows the shape of several neutrino spectra $\Psi^\odot_{e,j}(E)$.
The final neutrino energy spectrum  $\Psi^\odot_{e,j}(E)$  is significantly different from the original spectrum 
$\Psi^s_{e,j}(E)$. Indeed, while $\Psi^s_{e,j}(E)$  depends only on the properties of the nuclear reaction, 
$\Psi^\odot_{e,j}(E)$ becomes distinct from $\Psi^s_{e,j}(E)$ due 
to the contribution of neutrino flavour oscillations. Specifically, the $\Psi^\odot_{e,j}(E)$ depends on the fundamental 
parameters related with the neutrino vacuum oscillations  
through the (generalized) MSW oscillation 
mechanism, which depends on the local densities of electrons and quarks, 
and the $nsi-$ coupling constants~\citep{2013PhRvD..88d5006L}.
As discussed previously, all these effects are taken into account in $\langle P_{e} (E)\rangle_j $ (equation~\ref{eq:Pnuej}). 
In this study, we do not include the mono-energetic spectral lines of pp chains nuclear reactions $PeP$ and $^7Be$.
Although, the previous result (equation~\ref{eq-Psie}) also holds for these two neutrino sources
(corresponds to the sources marked with the subscript ($^*$) in figure~\ref{figure:1}), 
we opt for not including them in this study, since for these neutrino sources 
other solar plasma properties contribute to change the shape of the neutrino spectral lines. 

\smallskip
Figure~\ref{figure:6} shows the spectra of electron neutrinos for some of the leading nuclear reactions
of the Sun's core. The general shape of the spectra  $\Psi^\odot_{e,j}(E)$ (equation~\ref{eq-Psie}) 
is a combination of the neutrino spectrum of the nuclear reaction $\Psi^\odot_{e,j}(E)$,
and $\langle P_{e} (E)\rangle_j $ which depends of local density of electrons, down-quarks and up-quarks,
as well as of the $nsi$- parameters of the generalized MSW mechanism. For the specific set of $nsi$-parameters 
discussed in this study, clearly the  $HeP$ and $^8B$ neutrino emission shows the larger variation of the shape of their spectra. 

In both cases the interaction of neutrinos with (up- and down-) quarks leads to neutrino spectra 
with quite distinct shapes (blue and green areas in Figure~\ref{figure:6}) from the ones 
found in the standard MSW neutrino interaction (red area in Figure~\ref{figure:6}).
Equally important is the fact that it is possible to distinguish between the 
two neutrino models of interaction with quarks, since each model depends  differently of the neutrino energy.    
The $^{15}O$ and $^{17}F$ nuclear reactions also show neutrino spectra  with different shapes, 
although in this case the impact of the $nsi-$ interactions is much less pronounced than in the previous case,
at least for the current set of parameters. For the two other nuclear reactions, $PP$ and $^{13}N$, the impact of the $nsi-$ interactions is very small. This is somehow expected since the energy of the neutrinos emitted in these nuclear reactions is relatively small. For this neutrino energy range the  flavour  oscillations are dominated by vacuum oscillations  
and are almost independent of matter oscillations.   
 
\begin{figure}[!t]
\centering 
\includegraphics[scale=0.45]{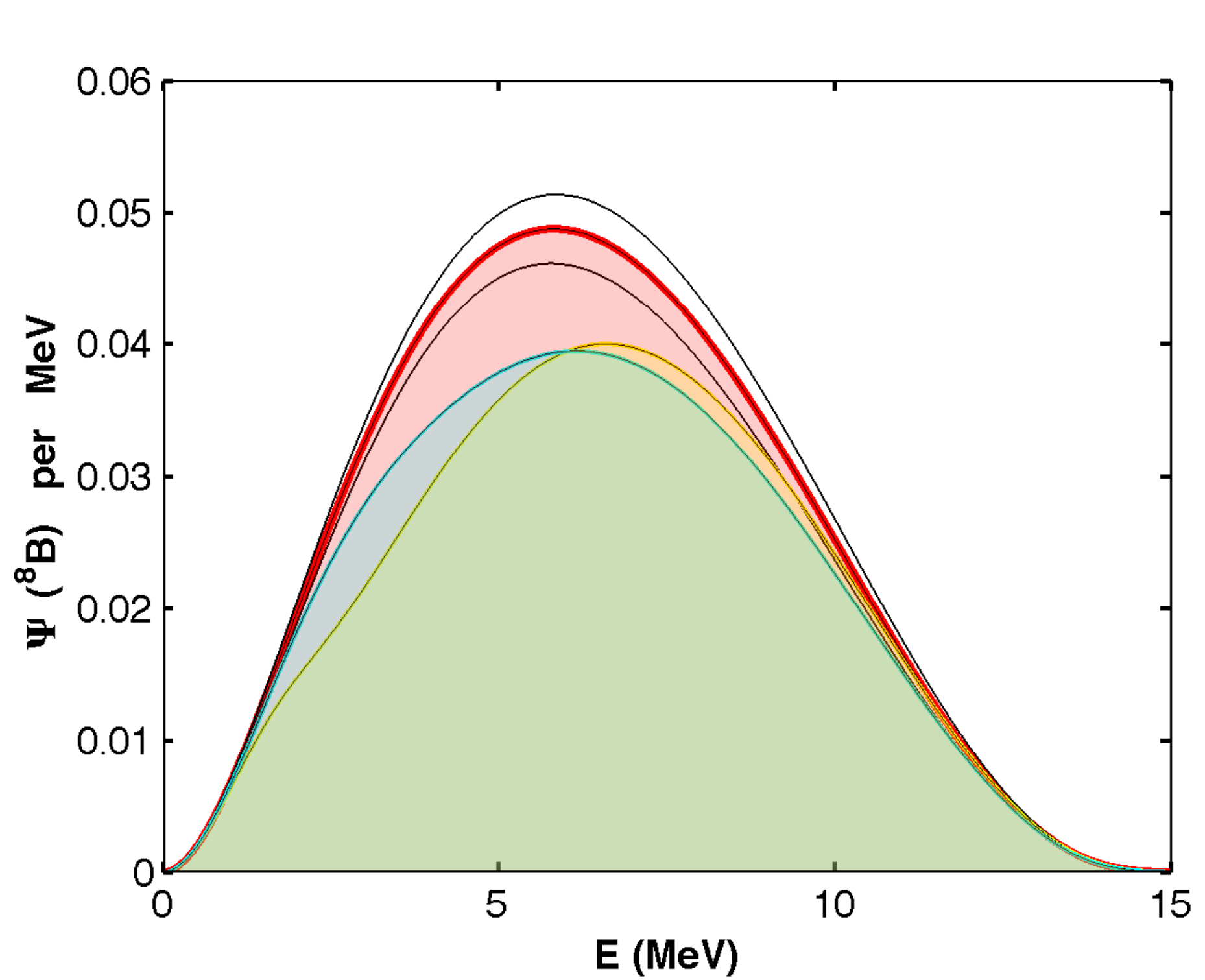}
\caption{
The $\Psi_{e,{^8B}}^{\odot} (E)$ is the electron solar neutrino spectrum 
for the standard  MSW effect (red area) with the error bar computed for the forthcoming LENA experiment. 
The error (black lines) in the spectrum shape is computed assuming the error in the survival probability is 
$P_{e}(E)\pm 0.025 $, which corresponds to 5 years of the LENA measurements.
The coloured curves follow the same colour scheme shown in Figure~\ref{figure:6}.
For clarity we have not included the error bar in the two other  curves. Nevertheless,
we note that the error bars for the other curves are identical to this one.}
\label{figure:7}
\end{figure}  
  
\section{Conclusion} 
\label{sec-DC}


In this study we have computed the expected alteration in the shape of some leading solar neutrino spectra resulting from neutrinos having a new type of interactions with up- and down-quarks, identical to the MSW oscillations of neutrinos with electrons. 
This new type of matter interaction, also known as generalized MSW oscillations,  
depends  of the specific properties of the neutrino interaction model 
but also of the local thermodynamic properties of the Sun's interior. 

The study shows that the neutrino spectra of the different solar nuclear reactions have quite 
distinct sensitivities to the new neutrino physics. The $HeP$ and $^8B$ 
neutrino spectra have their shapes  more affected by 
the new interaction between neutrinos and quarks. The $^{15}O$ and $^{17}F$ neutrino spectra 
also have a small alteration to their shapes, but these effects are much less pronounced than in the previous case. 
The impact of new physics in the  $PP$ and $^{13}N$ spectra is also very small. 

The new generation of neutrinos experiments such as Low Energy Neutrino Astronomy~\citep[LENA,][]{2012APh....35..685W}, 
Jiangmen Underground Neutrino Observatory~\citep[JUNO,][]{2016JPhG...43c0401A} 
Jinping Neutrino Experiment~\citep[Jinping,][]{2016arXiv160201733B},
Deep Underground Neutrino Experiment~\citep[DUNE,][]{2016NuPhB.908..318D},
and NO$\nu$A Neutrino Experiment~\citep[NO$\nu$A,][]{2012arXiv1207.6642F}, will allow to test some of new 
neutrino physics theories. The most promising evidence to discover $nsi-$ in solar neutrino data
is the precise measurement of the $^8B$ spectrum. Conveniently, we have estimated
how the experimental error of the next generation of detectors like LENA~\citep{2014PhLB..737..251M}
could affect our conclusion. In Figure~\ref{figure:7} is show an error bar estimation
on the $^8B$ spectrum computed assuming the error in the survival probability is $P_{e}(E)\pm 0.025 $, 
which is the precision possible to be obtained for the electron neutrino survival probability after 5 years 
of LENA measurements~\citep{2014PhLB..737..251M}. Even in a relatively short period of 5 years
of neutrino observations, it is already possible to find if neutrinos are experiencing 
flavour oscillations due to their interaction with quarks.       
Indeed, the identification by a future solar neutrino detector of a strong distortion in the shape of the solar neutrino
spectrum, like the $^8B$ neutrino spectrum  compared to the one predicted by the standard solar model,   
will  constitute  a strong indication for the existence of interactions between neutrinos and quarks in the Sun's core. 
The location and magnitude of the distortion of the solar spectrum  should give us some indication about the type of interaction (i.e., up- and down-quarks or both). 
 
In conclusion, we have shown that in the near future neutrino spectroscopic measurements will be used to infer 
the new interaction between neutrinos and quarks. This will be an important and totally independent way 
of testing new neutrino physics interaction models.

\begin{acknowledgments}
This work was supported by grants from Funda\c c\~ao para a Ci\^encia e Tecnologia  and Funda\c c\~ao Calouste Gulbenkian. The author thanks the anonymous referee for the  comments and suggestions made to this work.
\end{acknowledgments}

\appendix
\section{dimensionless parameters encoding the deviation from standard interactions} 
\label{sec-Ap}
The interactions of neutrinos with matter in the theoretical framework of the non-standard model
~\citep[e.g.,][]{2013JHEP...09..152G} can be described by the Lagrangian term given by equation~\ref{eq-Lagr_nsi}. 
In the following, it is assumed that electron neutrinos couples only with the up-quarks and down-quarks  
of the solar plasma (see section~\ref{sec-NNP} for details). For convenience,
we adopt the parametrization of~\citet{2004PhRvD..70k1301F,2013JHEP...09..152G,2016EPJA...52...87M}
in which the coupling of electron neutrinos with either up-quarks or down-quarks
of the solar plasma is parametrized by a set of two independent parameters ($\epsilon^u_N,\epsilon^u_D$) or
($\epsilon^d_N,\epsilon^d_D$).  Accordingly,  the coefficients $\epsilon^f_{D}$  and $\epsilon^f_{N}$
relate to the original parameters $\epsilon_{\alpha\beta}$ as
\begin{eqnarray}
\epsilon_{D}^f=c_{12}s_{13}
Re\left[e^{i\delta_{CP}}\left(s_{23}\epsilon_{e\mu}^f +c_{23}\epsilon_{e\tau}^f  \right) \right]
\nonumber
\\
-(1+s_{13}^2)c_{23}s_{23}Re\left(\epsilon_{\mu\tau}^f\right)
\nonumber
\\
-\frac{c^2_{13}}{2}\left(\epsilon_{ee}^f-\epsilon_{\mu\mu}^f\right)+\frac{s_{23}^2-s_{13}^2c_{23}^2}{2}
\left(\epsilon_{\tau\tau}^f -\epsilon_{\mu\mu}^f  \right) 
\end{eqnarray}
and
\begin{eqnarray}
\epsilon_{N}^f=c_{13}\left(c_{23}\epsilon_{e\mu}^f -s_{23}\epsilon_{e\tau}^f \right) 
\nonumber
\\
+s_{13}e^{-i\delta_{CP}} \left[s_{23}^2\epsilon_{\mu\tau}^f-c_{23}^2\epsilon_{\mu\tau}^{f*} 
+c_{23}s_{23} \left(\epsilon_{\tau\tau}^f -\epsilon_{\mu\mu}^f  \right)  \right].
\end{eqnarray}
As in this work we are only interested in the interaction of neutrinos with the solar plasma, 
we will consider at a time the following values of $f$: $e$, $u$ and $d$.
A detailed discussion about the relevance of this parametrization can be found
in~\citet[e.g.,][]{2004PhRvD..70k1301F,2013JHEP...09..152G}. 


\begin{thebibliography}{66}
\providecommand{\natexlab}[1]{#1}
\providecommand{\url}[1]{\texttt{#1}}
\expandafter\ifx\csname urlstyle\endcsname\relax
  \providecommand{\doi}[1]{doi: #1}\else
  \providecommand{\doi}{doi: \begingroup \urlstyle{rm}\Url}\fi

\bibitem[{Jung} et~al.(2001){Jung}, {Kajita}, {Mann}, and
  {McGrew}]{2001ARNPS..51..451J}
C.~K. {Jung}, T.~{Kajita}, T.~{Mann}, and C.~{McGrew}.
\newblock {Oscillations of Atmospheric Neutrinos}.
\newblock \emph{Annual Review of Nuclear and Particle Science}, 51:\penalty0
  451--488, 2001.
\newblock \doi{10.1146/annurev.nucl.51.101701.132421}.

\bibitem[{Kajita} and {Totsuka}(2001)]{2001RvMP...73...85K}
T.~{Kajita} and Y.~{Totsuka}.
\newblock {Observation of atmospheric neutrinos}.
\newblock \emph{Reviews of Modern Physics}, 73:\penalty0 85--118, January 2001.
\newblock \doi{10.1103/RevModPhys.73.85}.

\bibitem[{Ahmad}(2001)]{2001PhRvL..87g1301A}
Q.~R.~{\it et al.} {Ahmad}.
\newblock {Measurement of the Rate of Interactions $\nu_e$ + d $\rightarrow$ p
  + p + e - Produced by $^{8}$B Solar Neutrinos at the Sudbury Neutrino
  Observatory}.
\newblock \emph{Physical Review Letters}, 87\penalty0 (7):\penalty0 071301,
  August 2001.
\newblock \doi{10.1103/PhysRevLett.87.071301}.

\bibitem[{Davis} et~al.(1968){Davis}, {Harmer}, and
  {Hoffman}]{1968PhRvL..20.1205D}
R.~{Davis}, D.~S. {Harmer}, and K.~C. {Hoffman}.
\newblock {Search for Neutrinos from the Sun}.
\newblock \emph{Physical Review Letters}, 20:\penalty0 1205--1209, May 1968.
\newblock \doi{10.1103/PhysRevLett.20.1205}.

\bibitem[{Haxton} et~al.(2013){Haxton}, {Hamish Robertson}, and
  {Serenelli}]{2013ARAA..51...21H}
W.~C. {Haxton}, R.~G. {Hamish Robertson}, and A.~M. {Serenelli}.
\newblock {Solar Neutrinos: Status and Prospects}.
\newblock \emph{Annual Review of Astronomy and Astrophysics}, 51:\penalty0
  21--61, August 2013.
\newblock \doi{10.1146/annurev-astro-081811-125539}.

\bibitem[{Wurm}(2012)]{2012APh....35..685W}
M.~{\it et al.} {Wurm}.
\newblock {The next-generation liquid-scintillator neutrino observatory LENA}.
\newblock \emph{Astroparticle Physics}, 35:\penalty0 685--732, June 2012.
\newblock \doi{10.1016/j.astropartphys.2012.02.011}.

\bibitem[{An}(2016)]{2016JPhG...43c0401A}
F.~{\it et al.} {An}.
\newblock {Neutrino physics with JUNO}.
\newblock \emph{Journal of Physics G Nuclear Physics}, 43\penalty0
  (3):\penalty0 030401, March 2016.
\newblock \doi{10.1088/0954-3899/43/3/030401}.

\bibitem[{de Gouv{\^e}a} and {Kelly}(2016)]{2016NuPhB.908..318D}
A.~{de Gouv{\^e}a} and K.~J. {Kelly}.
\newblock {Non-standard neutrino interactions at DUNE}.
\newblock \emph{Nuclear Physics B}, 908:\penalty0 318--335, July 2016.
\newblock \doi{10.1016/j.nuclphysb.2016.03.013}.

\bibitem[{Friedland} and {Shoemaker}(2012)]{2012arXiv1207.6642F}
A.~{Friedland} and I.~M. {Shoemaker}.
\newblock {Searching for Novel Neutrino Interactions at NOvA and Beyond in
  Light of Large theta\_13}.
\newblock \emph{ArXiv e-prints:1207.6642}, July 2012.

\bibitem[{Beacom}(2016)]{2016arXiv160201733B}
J.~F.~{\it et al.} {Beacom}.
\newblock {Letter of Intent: Jinping Neutrino Experiment}.
\newblock \emph{ArXiv e-prints:1602.01733}, February 2016.

\bibitem[{Bandyopadhyay}(2009)]{2009RPPh...72j6201B}
A.~{\it et al.} {Bandyopadhyay}.
\newblock {Physics at a future Neutrino Factory and super-beam facility}.
\newblock \emph{Reports on Progress in Physics}, 72\penalty0 (10):\penalty0
  106201, October 2009.
\newblock \doi{10.1088/0034-4885/72/10/106201}.

\bibitem[{de Gouvea}(2013)]{2013arXiv1310.4340D}
A.~{\it et al.} {de Gouvea}.
\newblock {Neutrinos}.
\newblock \emph{ArXiv e-prints:1310.4340}, October 2013.

\bibitem[{Balantekin} and {Y{\"u}ksel}(2005)]{2005NuPhS.138..347B}
A.~B. {Balantekin} and H.~{Y{\"u}ksel}.
\newblock {Physics potential of solar neutrino experiments}.
\newblock \emph{Nuclear Physics B Proceedings Supplements}, 138:\penalty0
  347--349, January 2005.
\newblock \doi{10.1016/j.nuclphysbps.2004.11.080}.

\bibitem[{Balantekin} et~al.(1998){Balantekin}, {Beacom}, and
  {Fetter}]{1998PhLB..427..317B}
A.~B. {Balantekin}, J.~F. {Beacom}, and J.~M. {Fetter}.
\newblock {Matter-enhanced neutrino oscillations in the quasi-adiabatic limit}.
\newblock \emph{Physics Letters B}, 427:\penalty0 317--322, May 1998.
\newblock \doi{10.1016/S0370-2693(98)00231-7}.

\bibitem[{Lopes}(2013{\natexlab{a}})]{2013ApJ...777L...7L}
I.~{Lopes}.
\newblock {Search for Global f-Modes and p-Modes in the $^{8}$B Neutrino Flux}.
\newblock \emph{The Astrophysical Journal Letters}, 777:\penalty0 L7, November
  2013{\natexlab{a}}.
\newblock \doi{10.1088/2041-8205/777/1/L7}.

\bibitem[{Serenelli} et~al.(2013){Serenelli}, {Pe{\~n}a-Garay}, and
  {Haxton}]{2013PhRvD..87d3001S}
A.~{Serenelli}, C.~{Pe{\~n}a-Garay}, and W.~C. {Haxton}.
\newblock {Using the standard solar model to constrain solar composition and
  nuclear reaction S factors}.
\newblock \emph{\prd}, 87\penalty0 (4):\penalty0 043001, February 2013.
\newblock \doi{10.1103/PhysRevD.87.043001}.

\bibitem[{Lopes} and {Silk}(2002)]{2002PhRvL..88o1303L}
I.~P. {Lopes} and J.~{Silk}.
\newblock {Solar Neutrinos: Probing the Quasi-isothermal Solar Core Produced by
  Supersymmetric Dark Matter Particles}.
\newblock \emph{Physical Review Letters}, 88\penalty0 (15):\penalty0 151303,
  April 2002.
\newblock \doi{10.1103/PhysRevLett.88.151303}.

\bibitem[{Wendell}(2010)]{2010PhRvD..81i2004W}
R.~{\it et al.} {Wendell}.
\newblock {Atmospheric neutrino oscillation analysis with subleading effects in
  Super-Kamiokande I, II, and III}.
\newblock \emph{Physical Review D}, 81\penalty0 (9):\penalty0 092004, May 2010.
\newblock \doi{10.1103/PhysRevD.81.092004}.

\bibitem[{Adamson}(2013)]{2013PhRvL.110q1801A}
P.~{\it et al.} {Adamson}.
\newblock {Electron Neutrino and Antineutrino Appearance in the Full MINOS Data
  Sample}.
\newblock \emph{Physical Review Letters}, 110\penalty0 (17):\penalty0 171801,
  April 2013.
\newblock \doi{10.1103/PhysRevLett.110.171801}.

\bibitem[{Abe} et~al.(2014){Abe}, {Adam}, {Aihara}, {Akiri}, {Andreopoulos},
  {Aoki}, {Ariga}, {Ariga}, {Assylbekov}, {Autiero}, and
  et~al.]{2014PhRvL.112f1802A}
K.~{Abe}, J.~{Adam}, H.~{Aihara}, T.~{Akiri}, C.~{Andreopoulos}, S.~{Aoki},
  A.~{Ariga}, T.~{Ariga}, S.~{Assylbekov}, D.~{Autiero}, and et~al.
\newblock {Observation of Electron Neutrino Appearance in a Muon Neutrino
  Beam}.
\newblock \emph{Physical Review Letters}, 112\penalty0 (6):\penalty0 061802,
  February 2014.
\newblock \doi{10.1103/PhysRevLett.112.061802}.

\bibitem[{Miranda} and {Nunokawa}(2015)]{2015NJPh...17i5002M}
O.~G. {Miranda} and H.~{Nunokawa}.
\newblock {Non standard neutrino interactions: current status and future
  prospects}.
\newblock \emph{New Journal of Physics}, 17\penalty0 (9):\penalty0 095002,
  September 2015.
\newblock \doi{10.1088/1367-2630/17/9/095002}.

\bibitem[{Ohlsson}(2013)]{2013RPPh...76d4201O}
T.~{Ohlsson}.
\newblock {Status of non-standard neutrino interactions}.
\newblock \emph{Reports on Progress in Physics}, 76\penalty0 (4):\penalty0
  044201, April 2013.
\newblock \doi{10.1088/0034-4885/76/4/044201}.

\bibitem[{Girardi} et~al.(2014){Girardi}, {Meloni}, and
  {Petcov}]{2014NuPhB.886...31G}
I.~{Girardi}, D.~{Meloni}, and S.~T. {Petcov}.
\newblock {The Daya Bay and T2K results on $\sin^{2}\theta_{13}$ and
  non-standard neutrino interactions}.
\newblock \emph{Nuclear Physics B}, 886:\penalty0 31--42, September 2014.
\newblock \doi{10.1016/j.nuclphysb.2014.06.014}.

\bibitem[{Turck-Chieze} and {Lopes}(1993)]{1993ApJ...408..347T}
S.~{Turck-Chieze} and I.~{Lopes}.
\newblock {Toward a unified classical model of the sun - On the sensitivity of
  neutrinos and helioseismology to the microscopic physics}.
\newblock \emph{The Astrophysical Journal}, 408:\penalty0 347--367, May 1993.
\newblock \doi{10.1086/172592}.

\bibitem[{Morel}(1997)]{1997AAS..124..597M}
P.~{Morel}.
\newblock {CESAM: A code for stellar evolution calculations}.
\newblock \emph{A \& A Supplement series}, 124, 597, September 1997.
\newblock \doi{10.1051/aas:1997209}.

\bibitem[{Asplund} et~al.(2005){Asplund}, {Grevesse}, and
  {Sauval}]{2005ASPC..336...25A}
M.~{Asplund}, N.~{Grevesse}, and A.~J. {Sauval}.
\newblock {The Solar Chemical Composition}.
\newblock In T.~G. {Barnes}, III and F.~N. {Bash}, editors, \emph{Cosmic
  Abundances as Records of Stellar Evolution and Nucleosynthesis}, volume 336
  of \emph{Astronomical Society of the Pacific Conference Series, San Francisco}, page 25,   September 2005.

\bibitem[{Asplund} et~al.(2009){Asplund}, {Grevesse}, {Sauval}, and
  {Scott}]{2009ARAA..47..481A}
M.~{Asplund}, N.~{Grevesse}, A.~J. {Sauval}, and P.~{Scott}.
\newblock {The Chemical Composition of the Sun}.
\newblock \emph{Annual Review of Astronomy \& Astrophysics}, 47:\penalty0
  481--522, September 2009.
\newblock \doi{10.1146/annurev.astro.46.060407.145222}.

\bibitem[{Turck-Chi{\`e}ze} and {Couvidat}(2011)]{2011RPPh...74h6901T}
S.~{Turck-Chi{\`e}ze} and S.~{Couvidat}.
\newblock {Solar neutrinos, helioseismology and the solar internal dynamics}.
\newblock \emph{Reports on Progress in Physics}, 74\penalty0 (8):\penalty0
  086901, August 2011.
\newblock \doi{10.1088/0034-4885/74/8/086901}.

\bibitem[{Turck-Chi{\`e}ze} et~al.(2004){Turck-Chi{\`e}ze}, {Couvidat}, {Piau},
  {Ferguson}, {Lambert}, {Ballot}, {Garc{\'{\i}}a}, and
  {Nghiem}]{2004PhRvL..93u1102T}
S.~{Turck-Chi{\`e}ze}, S.~{Couvidat}, L.~{Piau}, J.~{Ferguson}, P.~{Lambert},
  J.~{Ballot}, R.~A. {Garc{\'{\i}}a}, and P.~{Nghiem}.
\newblock {Surprising Sun: A New Step Towards a Complete Picture?}
\newblock \emph{Physical Review Letters}, 93\penalty0 (21):\penalty0 211102,
  November 2004.
\newblock \doi{10.1103/PhysRevLett.93.211102}.

\bibitem[{Bahcall} et~al.(2005{\natexlab{a}}){Bahcall}, {Basu}, {Pinsonneault},
  and {Serenelli}]{2005ApJ...618.1049B}
J.~N. {Bahcall}, S.~{Basu}, M.~{Pinsonneault}, and A.~M. {Serenelli}.
\newblock {Helioseismological Implications of Recent Solar Abundance
  Determinations}.
\newblock \emph{The Astrophysical Journal}, 618:\penalty0 1049--1056, January
  2005{\natexlab{a}}.
\newblock \doi{10.1086/426070}.

\bibitem[{Bahcall} et~al.(2005{\natexlab{b}}){Bahcall}, {Serenelli}, and
  {Basu}]{2005ApJ...621L..85B}
J.~N. {Bahcall}, A.~M. {Serenelli}, and S.~{Basu}.
\newblock {New Solar Opacities, Abundances, Helioseismology, and Neutrino
  Fluxes}.
\newblock \emph{The Astrophysical Journal}, 621:\penalty0 L85--L88, March
  2005{\natexlab{b}}.
\newblock \doi{10.1086/428929}.

\bibitem[{Serenelli} et~al.(2009){Serenelli}, {Basu}, {Ferguson}, and
  {Asplund}]{2009ApJ...705L.123S}
A.~M. {Serenelli}, S.~{Basu}, J.~W. {Ferguson}, and M.~{Asplund}.
\newblock {New Solar Composition: The Problem with Solar Models Revisited}.
\newblock \emph{The Astrophysical Journal Letters}, 705:\penalty0 L123--L127,
  November 2009.
\newblock \doi{10.1088/0004-637X/705/2/L123}.

\bibitem[{Guzik} and {Mussack}(2010)]{2010ApJ...713.1108G}
J.~A. {Guzik} and K.~{Mussack}.
\newblock {Exploring Mass Loss, Low-Z Accretion, and Convective Overshoot in
  Solar Models to Mitigate the Solar Abundance Problem}.
\newblock \emph{The Astrophysical Journal}, 713:\penalty0 1108--1119, April
  2010.
\newblock \doi{10.1088/0004-637X/713/2/1108}.

\bibitem[{Turck-Chi{\`e}ze} et~al.(2010){Turck-Chi{\`e}ze}, {Palacios},
  {Marques}, and {Nghiem}]{2010ApJ...715.1539T}
S.~{Turck-Chi{\`e}ze}, A.~{Palacios}, J.~P. {Marques}, and P.~A.~P. {Nghiem}.
\newblock {Seismic and Dynamical Solar Models. I. The Impact of the Solar
  Rotation History on Neutrinos and Seismic Indicators}.
\newblock \emph{The Astrophysical Journal}, 715:\penalty0 1539--1555, June
  2010.
\newblock \doi{10.1088/0004-637X/715/2/1539}.

\bibitem[{Turck-Chi{\`e}ze} and {Lopes}(2012)]{2012RAA....12.1107T}
S.~{Turck-Chi{\`e}ze} and I.~{Lopes}.
\newblock {Solar-stellar astrophysics and dark matter}.
\newblock \emph{Research in Astronomy and Astrophysics}, 12:\penalty0
  1107--1138, August 2012.
\newblock \doi{10.1088/1674-4527/12/8/011}.

\bibitem[{Grevesse} and {Sauval}(1998)]{1998SSRv...85..161G}
N.~{Grevesse} and A.~J. {Sauval}.
\newblock {Standard Solar Composition}.
\newblock \emph{Space Science Reviews}, 85:\penalty0 161--174, May 1998.
\newblock \doi{10.1023/A:1005161325181}.

\bibitem[{Vagnozzi} et~al.(2016){Vagnozzi}, {Freese}, and
  {Zurbuchen}]{2016arXiv160305960V}
S.~{Vagnozzi}, K.~{Freese}, and T.~H. {Zurbuchen}.
\newblock {A successful solar model using new solar composition data}.
\newblock \emph{ArXiv e-prints:1603.05960}, March 2016.

\bibitem[{Gonzalez-Garcia} and {Maltoni}(2008)]{2008PhR...460....1G}
M.~C. {Gonzalez-Garcia} and M.~{Maltoni}.
\newblock {Phenomenology with massive neutrinos}.
\newblock \emph{Physics Reports}, 460:\penalty0 1--129, April 2008.
\newblock \doi{10.1016/j.physrep.2007.12.004}.

\bibitem[{Bellini} et~al.(2013){Bellini}, {Ludhova}, {Ranucci}, and
  {Villante}]{2013arXiv1310.7858B}
G.~{Bellini}, L.~{Ludhova}, G.~{Ranucci}, and F.~L. {Villante}.
\newblock {Neutrino oscillations.}
\newblock \emph{Advances in High Energy Physics}, Volume 2014 (2014),
\newblock {Article ID 191960},
\newblock \doi{10.1155/2014/191960}.

\bibitem[{Papoulias} and {Kosmas}(2015)]{2015arXiv150202928P}
D.~K. {Papoulias} and T.~S. {Kosmas}.
\newblock {Standard and non-standard neutrino-nucleus reactions cross sections   and event rates to neutrino detection experiments}.
\newblock \emph{Advances in High Energy Physics}, Volume 2015 (2015),
\newblock {Article ID 763648},
\newblock \doi{10.1155/2015/763648}.

\bibitem[{Davidson} et~al.(2003){Davidson}, {na-Garay}, {Rius}, and
  {Santamaria}]{2003JHEP...03..011D}
S.~{Davidson}, C.~P. {na-Garay}, N.~{Rius}, and A.~{Santamaria}.
\newblock {Present and future bounds on non-standard neutrino interactions}.
\newblock \emph{Journal of High Energy Physics}, 3:\penalty0 011, March 2003.
\newblock \doi{10.1088/1126-6708/2003/03/011}.

\bibitem[{Donnelly} and {Walecka}(1976)]{1976NuPhA.274..368D}
T.~W. {Donnelly} and J.~D. {Walecka}.
\newblock {Semi-leptonic weak and electromagnetic interactions with nuclei:
  Isoelastic processes}.
\newblock \emph{Nuclear Physics A}, 274:\penalty0 368--412, December 1976.
\newblock \doi{10.1016/0375-9474(76)90209-8}.

\bibitem[{N{\"o}tzold} and {Raffelt}(1988)]{1988NuPhB.307..924N}
D.~{N{\"o}tzold} and G.~{Raffelt}.
\newblock {Neutrono dispersion at finite temperature and density}.
\newblock \emph{Nuclear Physics B}, 307:\penalty0 924--936, October 1988.
\newblock \doi{10.1016/0550-3213(88)90113-7}.

\bibitem[{Olive} and {Particle Data Group}(2014)]{2014ChPhC..38i0001O}
K.~A. {Olive} and {Particle Data Group}.
\newblock {Review of Particle Physics}.
\newblock \emph{Chinese Physics C}, 38\penalty0 (9):\penalty0 090001, August
  2014.
\newblock \doi{10.1088/1674-1137/38/9/090001}.

\bibitem[{Giunti} and {Chung}(2007)]{2007fnpa.book.....G}
C.~{Giunti} and W.~K. {Chung}.
\newblock \emph{{Fundamentals of Neutrino Physics and Astrophysics}}.
\newblock Oxford University Press, 2007.

\bibitem[{Kuo} and {Pantaleone}(1989)]{1989RvMP...61..937K}
T.~K. {Kuo} and J.~{Pantaleone}.
\newblock {Neutrino oscillations in matter}.
\newblock \emph{Reviews of Modern Physics}, 61:\penalty0 937--980, October
  1989.
\newblock \doi{10.1103/RevModPhys.61.937}.

\bibitem[{Huitu} et~al.(2016){Huitu}, {K{\"a}rkk{\"a}inen}, {Maalampi}, and
  {Vihonen}]{2016PhRvD..93e3016H}
K.~{Huitu}, T.~J. {K{\"a}rkk{\"a}inen}, J.~{Maalampi}, and S.~{Vihonen}.
\newblock {Constraining the nonstandard interaction parameters in long baseline
  neutrino experiments}.
\newblock \emph{\prd}, 93\penalty0 (5):\penalty0 053016, March 2016.
\newblock \doi{10.1103/PhysRevD.93.053016}.

\bibitem[{Bergstr{\"o}m} et~al.(2015){Bergstr{\"o}m}, {Gonzalez-Garcia},
  {Maltoni}, and {Schwetz}]{2015JHEP...09..200B}
J.~{Bergstr{\"o}m}, M.~C. {Gonzalez-Garcia}, M.~{Maltoni}, and T.~{Schwetz}.
\newblock {Bayesian global analysis of neutrino oscillation data}.
\newblock \emph{Journal of High Energy Physics}, 9:\penalty0 200, September
  2015.
\newblock \doi{10.1007/JHEP09(2015)200}.

\bibitem[{Gonzalez-Garcia} et~al.(2014){Gonzalez-Garcia}, {Maltoni}, and
  {Schwetz}]{2014JHEP...11..052G}
M.~C. {Gonzalez-Garcia}, M.~{Maltoni}, and T.~{Schwetz}.
\newblock {Updated fit to three neutrino mixing: status of leptonic CP
  violation}.
\newblock \emph{Journal of High Energy Physics}, 11:\penalty0 52, November
  2014.
\newblock \doi{10.1007/JHEP11(2014)052}.

\bibitem[{Apollonio} and {Baldini}(1999)]{1999PhLB..466..415A}
M.~{Apollonio} and A.~{Baldini}.
\newblock {Limits on neutrino oscillations from the CHOOZ experiment}.
\newblock \emph{Physics Letters B}, 466:\penalty0 415--430, November 1999.
\newblock \doi{10.1016/S0370-2693(99)01072-2}.

\bibitem[{Fogli} et~al.(2009){Fogli}, {Lisi}, {Marrone}, {Palazzo}, and
  {Rotunno}]{2009arXiv0905.3549F}
G.~L. {Fogli}, E.~{Lisi}, A.~{Marrone}, A.~{Palazzo}, and A.~M. {Rotunno}.
\newblock {Neutrino oscillations, global analysis and theta(13)}.
\newblock \emph{ArXiv e-prints:0905.3549}, May 2009.

\bibitem[{Fogli} et~al.(2012){Fogli}, {Lisi}, {Marrone}, {Montanino},
  {Palazzo}, and {Rotunno}]{2012PhRvD..86a3012F}
G.~L. {Fogli}, E.~{Lisi}, A.~{Marrone}, D.~{Montanino}, A.~{Palazzo}, and A.~M.
  {Rotunno}.
\newblock {Global analysis of neutrino masses, mixings, and phases: Entering
  the era of leptonic CP violation searches}.
\newblock \emph{Physical Review D}, 86\penalty0 (1):\penalty0 013012, July
  2012.
\newblock \doi{10.1103/PhysRevD.86.013012}.

\bibitem[{Hernandez}(2010)]{2010arXiv1010.4131H}
P.~{Hernandez}.
\newblock {Neutrino physics}.
\newblock \emph{ArXiv e-prints:1010.4131}, October 2010.

\bibitem[{Balantekin} and {Yuksel}(2003)]{2003hep.ph....1072B}
A.~B. {Balantekin} and H.~{Yuksel}.
\newblock {Global Analysis of Solar Neutrino and KamLAND Data}.
\newblock \emph{ArXiv High Energy Physics - Phenomenology e-prints:hep-ph/0301072}, January 2003.

\bibitem[{Goswami} and {Smirnov}(2005)]{2005PhRvD..72e3011G}
S.~{Goswami} and A.~Y. {Smirnov}.
\newblock {Solar neutrinos and 1-3 leptonic mixing}.
\newblock \emph{Physical Review D}, 72\penalty0 (5):\penalty0 053011, September
  2005.
\newblock \doi{10.1103/PhysRevD.72.053011}.

\bibitem[{Thomson}(2013)]{2013mpp..book.....T}
M.~{Thomson}.
\newblock \emph{{Modern Particle Physics}}.
\newblock (Cambridge University Press. Cambridge, UK, 2013)

\bibitem[{Gonzalez-Garcia} and {Maltoni}(2013)]{2013JHEP...09..152G}
M.~C. {Gonzalez-Garcia} and M.~{Maltoni}.
\newblock {Determination of matter potential from global analysis of neutrino
  oscillation data}.
\newblock \emph{Journal of High Energy Physics}, 9:\penalty0 152, September
  2013.
\newblock \doi{10.1007/JHEP09(2013)152}.

\bibitem[{Farzan}(2015)]{2015PhLB..748..311F}
Y.~{Farzan}.
\newblock {A model for large non-standard interactions of neutrinos leading to
  the LMA-Dark solution}.
\newblock \emph{Physics Letters B}, 748:\penalty0 311--315, September 2015.
\newblock \doi{10.1016/j.physletb.2015.07.015}.

\bibitem[{Maltoni} and {Smirnov}(2016)]{2016EPJA...52...87M}
M.~{Maltoni} and A.~Y. {Smirnov}.
\newblock {Solar neutrinos and neutrino physics}.
\newblock \emph{European Physical Journal A}, 52:\penalty0 87, April 2016.
\newblock \doi{10.1140/epja/i2016-16087-0}.

\bibitem[{Friedland} et~al.(2004){Friedland}, {Lunardini}, and
  {Maltoni}]{2004PhRvD..70k1301F}
A.~{Friedland}, C.~{Lunardini}, and M.~{Maltoni}.
\newblock {Atmospheric neutrinos as probes of neutrino-matter interactions}.
\newblock \emph{Physical Review D,}, 70\penalty0 (11):\penalty0 111301,
  December 2004.
\newblock \doi{10.1103/PhysRevD.70.111301}.

\bibitem[{Ortiz} et~al.(2000){Ortiz}, {Garc{\'{\i}}a}, {Waltz}, {Bhattacharya},
  and {Komives}]{2000PhRvL..85.2909O}
C.~E. {Ortiz}, A.~{Garc{\'{\i}}a}, R.~A. {Waltz}, M.~{Bhattacharya}, and A.~K.
  {Komives}.
\newblock {Shape of the $^{8}$B Alpha and Neutrino Spectra}.
\newblock \emph{Physical Review Letters}, 85:\penalty0 2909--2912, October
  2000.
\newblock \doi{10.1103/PhysRevLett.85.2909}.

\bibitem[{Winter} et~al.(2006){Winter}, {Freedman}, {Rehm}, and
  {Schiffer}]{2006PhRvC..73b5503W}
W.~T. {Winter}, S.~J. {Freedman}, K.~E. {Rehm}, and J.~P. {Schiffer}.
\newblock {The $^{8}$B neutrino spectrum}.
\newblock \emph{Physical Review C}, 73\penalty0 (2):\penalty0 025503, February
  2006.
\newblock \doi{10.1103/PhysRevC.73.025503}.

\bibitem[{Bahcall} and {Holstein}(1986)]{1986PhRvC..33.2121B}
J.~N. {Bahcall} and B.~R. {Holstein}.
\newblock {Solar neutrinos from the decay of $^{8}$B}.
\newblock \emph{Physical Review C}, 33:\penalty0 2121--2127, June 1986.
\newblock \doi{10.1103/PhysRevC.33.2121}.

\bibitem[{Napolitano} et~al.(1987){Napolitano}, {Freedman}, and
  {Camp}]{1987PhRvC..36..298N}
J.~{Napolitano}, S.~J. {Freedman}, and J.~{Camp}.
\newblock {Beta and neutrino spectra in the decay of $^{8}$B}.
\newblock \emph{Physical Review C}, 36:\penalty0 298--302, July 1987.
\newblock \doi{10.1103/PhysRevC.36.298}.

\bibitem[{Winter} et~al.(2003){Winter}, {Freedman}, {Rehm}, {Ahmad}, {Greene},
  {Heinz}, {Henderson}, {Janssens}, {Jiang}, {Moore}, {Mukherjee}, {Pardo},
  {Pennington}, {Savard}, {Schiffer}, {Seweryniak}, {Zinkann}, and
  {Paul}]{2003PhRvL..91y2501W}
W.~T. {Winter}, S.~J. {Freedman}, K.~E. {Rehm}, I.~{Ahmad}, J.~P. {Greene},
  A.~{Heinz}, D.~{Henderson}, R.~V. {Janssens}, C.~L. {Jiang}, E.~F. {Moore},
  G.~{Mukherjee}, R.~C. {Pardo}, T.~{Pennington}, G.~{Savard}, J.~P.
  {Schiffer}, D.~{Seweryniak}, G.~{Zinkann}, and M.~{Paul}.
\newblock {Determination of the $^{8}$B Neutrino Spectrum}.
\newblock \emph{Physical Review Letters}, 91\penalty0 (25):\penalty0 252501,
  December 2003.
\newblock \doi{10.1103/PhysRevLett.91.252501}.

\bibitem[{Roger}(2012)]{2012PhRvL.108p2502R}
T.~{\it et al.} {Roger}.
\newblock {Precise Determination of the Unperturbed B8 Neutrino Spectrum}.
\newblock \emph{Physical Review Letters}, 108\penalty0 (16):\penalty0 162502,
  April 2012.
\newblock \doi{10.1103/PhysRevLett.108.162502}.

\bibitem[{Bhattacharya} et~al.(2006){Bhattacharya}, {Adelberger}, and
  {Swanson}]{2006PhRvC..73e5802B}
M.~{Bhattacharya}, E.~G. {Adelberger}, and H.~E. {Swanson}.
\newblock {Precise study of the final-state continua in Li8 and B8 decays}.
\newblock \emph{Physical Review C}, 73\penalty0 (5):\penalty0 055802, May 2006.
\newblock \doi{10.1103/PhysRevC.73.055802}.

\bibitem[{Kirsebom} et~al.(2011){Kirsebom}, {Hyldegaard}, {Alcorta}, {Borge},
  {B{\"u}scher}, {Eronen}, {Fox}, {Fulton}, {Fynbo}, {Hultgren}, {Jokinen},
  {Jonson}, {Kankainen}, {Karvonen}, {Kessler}, {Laird}, {Madurga}, {Moore},
  {Nyman}, {Penttil{\"a}}, {Rahaman}, {Reponen}, {Riisager}, {Roger},
  {Ronkainen}, {Saastamoinen}, {Tengblad}, and
  {{\"A}yst{\"o}}]{2011PhRvC..83f5802K}
O.~S. {Kirsebom}, S.~{Hyldegaard}, M.~{Alcorta}, M.~J.~G. {Borge},
  J.~{B{\"u}scher}, T.~{Eronen}, S.~{Fox}, B.~R. {Fulton}, H.~O.~U. {Fynbo},
  H.~{Hultgren}, A.~{Jokinen}, B.~{Jonson}, A.~{Kankainen}, P.~{Karvonen},
  T.~{Kessler}, A.~{Laird}, M.~{Madurga}, I.~{Moore}, G.~{Nyman},
  H.~{Penttil{\"a}}, S.~{Rahaman}, M.~{Reponen}, K.~{Riisager}, T.~{Roger},
  J.~{Ronkainen}, A.~{Saastamoinen}, O.~{Tengblad}, and J.~{{\"A}yst{\"o}}.
\newblock {Precise and accurate determination of the B8 decay spectrum}.
\newblock \emph{Physical Review C}, 83\penalty0 (6):\penalty0 065802, June
  2011.
\newblock \doi{10.1103/PhysRevC.83.065802}.

\bibitem[{Lopes}(2013{\natexlab{b}})]{2013PhRvD..88d5006L}
I.~{Lopes}.
\newblock {Probing the Sun's inner core using solar neutrinos: A new diagnostic
  method}.
\newblock \emph{Physical Review D}, 88\penalty0 (4):\penalty0 045006, August
  2013{\natexlab{b}}.
\newblock \doi{10.1103/PhysRevD.88.045006}.

\bibitem[{M{\"o}llenberg} et~al.(2014){M{\"o}llenberg}, {von Feilitzsch},
  {Hellgartner}, {Oberauer}, {Tippmann}, {Winter}, {Wurm}, and
  {Zimmer}]{2014PhLB..737..251M}
R.~{M{\"o}llenberg}, F.~{von Feilitzsch}, D.~{Hellgartner}, L.~{Oberauer},
  M.~{Tippmann}, J.~{Winter}, M.~{Wurm}, and V.~{Zimmer}.
\newblock {Detecting the upturn of the solar $^{8}$B neutrino spectrum with
  LENA}.
\newblock \emph{Physics Letters B}, 737:\penalty0 251--255, October 2014.
\newblock \doi{10.1016/j.physletb.2014.08.053}.

\end{thebibliography}
\bibliographystyle{yahapj}

\end{document}